\documentclass{article}
\usepackage[utf8]{inputenc}
\usepackage{authblk}
\usepackage{setspace}
\usepackage[margin=1.25in]{geometry}
\usepackage{graphicx}
\graphicspath{ {./figures/} }
\usepackage{subcaption}
\usepackage{amsmath}
\usepackage{hyperref}
\usepackage{xcolor}
\usepackage{latexsym}
\usepackage{amssymb}
\usepackage{amsfonts}
\usepackage{amsthm}
\usepackage{bbm}
\usepackage{color}
\usepackage{cancel} 
\usepackage{ulem}
\usepackage{bm}
\usepackage{soul}
\usepackage{array}
\usepackage{multirow}
\usepackage[all]{xy}
\usepackage[bottom]{footmisc}
\usepackage{svg}

\hypersetup{
	bookmarksnumbered,
	unicode,			
	colorlinks,			
	citecolor=[rgb]{.9,0,.5},	
	urlcolor=[rgb]{0,0,1},	
	linkcolor=[rgb]{0,.7,0}	
	}

\usepackage[
    natbib=true,
    style=numeric-comp,
    sorting=none
]{biblatex}

\catcode`@=11
\def\secteqno{\@addtoreset{equation}{section}%
	\def\theequation{\thesection.\arabic{equation}}}
\catcode`@=12
\secteqno
\def\dd{\hbox{\,\Large$\triangleright$}}
\newcommand{\be}{\begin{equation}}
	\newcommand{\ee}{\end{equation}}
\newcommand{\bea}{\begin{eqnarray}}
	\newcommand{\eea}{\end{eqnarray}}
\newcommand{\bref}[1]{(\ref{#1})}
\newcommand{\nn}{\nonumber}

\def\dig#1{\setbox0=\hbox{$#1M$}
	\hskip.06\wd0 \vrule width.07\wd0 height.63\wd0 depth.01\wd0 
	\vrule width.37\wd0 height.63\wd0 depth-.56\wd0 \hskip-.4\wd0
	\vrule width.25\wd0 height.35\wd0 depth-.28\wd0 
	\vrule width.07\wd0 height.35\wd0 depth-.17\wd0 \hskip.14\wd0}
\def\digamma{{\mathpalette\dig{}}}
\def\={{\;=\;}}\def\+{{\;+\;}}

\newcommand{\B}{\beta}
\def\dig#1{\setbox0=\hbox{$#1M$}
	\hskip.06\wd0 \vrule width.08\wd0 height.63\wd0 depth.01\wd0 
	\vrule width.37\wd0 height.63\wd0 depth-.55\wd0 \hskip-.4\wd0
	\vrule width.25\wd0 height.36\wd0 depth-.28\wd0 
	\vrule width.08\wd0 height.36\wd0 depth-.17\wd0 \hskip.14\wd0}
\def\digamma{{\mathpalette\dig{}}}
\def\bop#1{\setbox0=\hbox{$#1M$}\mkern1.5mu
	\vbox{\hrule height0pt depth.1\ht0
		\hbox{\vrule width.1\ht0 height.8\ht0 \kern.8\ht0
			\vrule width.1\ht0}\hrule height.1\ht0}\mkern1.5mu}

\def\dd{\hbox{\,\Large$\triangleright$}} 

\def\don#1#2{{\buildrel{\mkern2.5mu\raise-.1em\hbox{$\scriptstyle#1$}\mkern-2.5mu}\over{#2}}}	
\def\dron#1#2{{\buildrel{{\raise-.1em\hbox{$\scriptstyle#1$}}}\over{#2}}}		
\newcommand{\ctext}[1]{\raise0.2ex\hbox{\textcircled{\scriptsize{#1}}}}

\title{Strings and membranes from ${\cal A}$-theory five brane }

\author[$\heartsuit$$\diamondsuit$]{Machiko Hatsuda\footnote{Email: \href{mhatsuda@juntendo.ac.jp}{mhatsuda@juntendo.ac.jp} }}
\author[$\ddagger$]{Ond\v{r}ej Hul\'{\i}k}
\author[$\#$]{ William D. Linch }
\author[$\clubsuit$$\spadesuit$]{ Warren D. Siegel }
\author[$\clubsuit$]{\\Di Wang}
\author[$\clubsuit$]{ Yu-Ping Wang 
}

\affil[$\clubsuit$]{\textit{ Department of Physics, SUNY Stony Brook University, Stony Brook, NY 11794, USA}}
\affil[$\spadesuit$]{\textit{C. N. Yang Institute for Theoretical Physics. Stony Brook, NY 11794, USA}}
\affil[$\heartsuit$]{\textit{Department of Radiological Technology, Faculty of Health Science, Juntendo University
		Yushima, Bunkyou-ku, Tokyo 113-0034, Japan}}
\affil[$\diamondsuit$]{\textit{KEK Theory Center, High Energy Accelerator Research Organization
		Tsukuba, Ibaraki 305-0801, Japan}}
\affil[$\ddagger$]{\textit{Institute for Mathematics 
 Ruprecht-Karls-Universitat Heidelberg,
 69120 Heidelberg, Germany}}
\affil[$\#$]{\textit{Thomas Jefferson High School for Science and Technology, Alexandria, VA 22312, USA} }
\date{\today}

\begin{document}
	
	\newgeometry{top=0.1in,bottom=1in,right=1in,left=1in}
	\hfill YITP-SB-2024-21
	\vskip 0.1in
	\hfill KEK-TH-2656
	{\let\newpage\relax\maketitle}
	
\begin{abstract}
The ${\cal A}$-theory takes U-duality symmetry as a guiding principle, with the SL(5) U-duality symmetry being described as the world-volume theory of a 5-brane. Furthermore, by unifying the 6-dimensional world-volume Lorentz symmetry with the SL(5) spacetime symmetry, it extends to SL(6) U-duality symmetry. 
The SL(5) spacetime vielbein fields and the 5-brane world-volume vielbein fields are mixed under the SL(6) U-duality transformation.
We demonstrate that consistent sectionings of the SL(6) ${\cal A}$5-brane world-volume Lagrangian yield 
Lagrangians of 
the ${\cal T}$-string with O(D,D) T-duality symmetry, the conventional string, the ${\cal M}$5-brane with GL(4) duality symmetry, and the non-perturbative M2-brane in supergravity theory. 
The GL(4) covariant Lagrangian of the ${\cal M}$5-brane derived in this manner is a new, perturbatively quantizable theory.

\end{abstract}
	
\restoregeometry
\tableofcontents
\newpage
\section{Introduction}


\subsection{An overview of \texorpdfstring{$\cal A$}{A}-theory formalism}

Superstring theory is regarded as a strong candidate for a unified theory that encompasses all four fundamental interactions. However, rather than a single theory, it currently exists in the form of six distinct theories: five superstring theories and M-theory. These six theories are intricately connected via S-duality and T-duality, forming a hexagonal network of dualities. Each of these six theories is defined on its own characteristic brane, and together they provide complementary descriptions of the broader structure of superstring theory. A key open question is what kind of theoretical framework can provide a unified formulation of these six theories.

S-duality and T-duality are unified as U-duality, whose structure reflects the group-theoretic inclusiveness of these dualities. The symmetry group of a theory with manifest S-duality is given by GL(D+1) which extends the spacetime diffeomorphism group GL(D) by incorporating the S-duality group SL(2) as a subgroup. GL(D+1) and the T-duality symmetry O(D,D) are embedded within the exceptional group E$_{\rm D+1}$ which serves as the symmetry group of U-duality. U-duality is therefore expected to relate  all six superstring theories in a coherent manner. A theory that explicitly manifests this U-duality symmetry is referred to as "${\cal A}$-theory," and major objective is to construct a perturbative formulation of this theory that facilitates a systematic analysis of its quantum aspects (see  \cite{Hatsuda:2023dwx} for review).

In 1995, Witten proposed M-theory as a strong coupling limit of the type IIA superstring theory via S-duality \cite{Witten:1995ex}. The low-energy effective theory of M-theory is 11-dimensional supergravity, whose diffeomorphism symmetry is GL(11), combining the 10-dimensional diffeomorphism and the SL(2) S-duality symmetry. We refer to a world-volume theories that exhibit manifest GL(D+1) symmetry as ${\cal M}$-theory. 
In 1996, Vafa proposed F-theory as a framework related to the type IIB superstring theory via S-duality \cite{Vafa:1996xn}. For further discussion, see, for example \cite{Weigand:2018rez}.

In 1993, Siegel proposed a string theory guided by the T-duality symmetry O(D,D) \cite{Siegel:1993th, Siegel:1993xq, Siegel:1993bj}. This framework employs the idea of doubling spacetime coordinates, where the 2D-dimensional spacetime coordinates are treated as vectors in the O(D,D) representation\cite{Tseytlin:1990va}. 
The geometry generated by the O(D,D) covariant current algebra is the stringy gravity theory with $B$-field, 
where gauge fields are parameters of the coset O(D,D) over the doubled Lorentz group. We refer to this as ${\cal T}$-theory and the T-duality covariant string as ${\cal T}$-string. Later Hull, Zwiebach, and Hohm proposed a theory of O(D,D) covariant background fields, known as Double Field Theory (DFT) \cite{Hull:2009mi,
Hohm:2010pp}. This theory serves as the low-energy effective gravitational description of ${\cal T}$-string theory. In order to consistently reduce the doubled 2D-dimensional spacetime to the physical D-dimensional spacetime, the section condition was employed, which corresponds to the zero-mode of the Virasoro constraint  ${\cal S}$=0. For detailed reviews, see \cite{Aldazabal:2013sca,Hohm:2013bwa}.
 The ${\cal T}$- theory was extended to incorporate $\mathcal{N}=2$ supersymmetry including Ramond–Ramond gauge fields based on the doubled non-degenerate super-Poincaré group \cite{Polacek:2014cva, Polacek:2016nry, Hatsuda:2014qqa, Hatsuda:2014aza, Hatsuda:2015cia}. All bosonic component fields represent supersymmetrized O(D,D) while fermions represent only O(D-1,1)$^2$.
Incorporating S-duality requires extending the symmetry to the exceptional U-duality group.  Consequently, the exceptional group E$_{\rm D+1}$ is expected to be related to a subgroup of the doubled non-degenerate super-Poincaré group.

Siegel, Linch and Polacek subsequently proposed  brane theories guided by the U-duality symmetry ( E$_{\rm D+1}$ ) \cite{Polacek:2014cva, Linch:2015fya, Linch:2015qva}. 
Since the representation of exceptional groups varies with dimension, the theory is labeled by the spacetime dimension $D$ when reduced to string theory. This is called D$=D$
${\cal A}$-theory. According to the classification of Lie algebras, removing a specific node in the Dynkin diagram of  E$_{\rm D+1}$ reduces it to the  GL(D+1)  Dynkin diagram. This corresponds to reducing the spacetime dimensions of ${\cal A}$-theory using the Virasoro constraint, recovering the aforementioned ${\cal M}$-theory. Conversely, removing another node in the E$_{\rm D+1}$ Dynkin diagram reduces it to the O(D,D) Dynkin diagram. This corresponds to reducing the world-volume dimensions of ${\cal A}$-theory using the Gauss law constraint ${\cal U}$=0, recovering the aforementioned ${\cal T}$-theory. As a consequence of these dual reductions, ${\cal A}$-theory is consistently described by branes covariant under the exceptional group. Both the spacetime coordinates and world-volume coordinates of the branes are representations of the exceptional group, ensuring that the brane current algebra is covariant under the exceptional group. Moreover, a perturbative Lagrangian describing these branes on their world-volume has been constructed. 
In the previous papers \cite{Linch:2015fya, Linch:2015qva, Linch:2016ipx, Linch:2015fca, Siegel:2018puf, Linch:2015lwa, Siegel:2019wrr, Hatsuda:2021wpb, Hatsuda:2021ezo, Siegel:2016dek, Siegel:2020gro, Ju:2016hla, Siegel:2020qef, Siegel:2019wrr, Linch:2017eru} we refer these models as ``F-theory'', but 
we have renamed our formulation from "F-theory" to ${\cal A}$-theory in our most recent work \cite{Hatsuda:2023dwx}, in order to present it as a general framework that includes all spacetime dimensions and accommodates all duality symmetries.

Exceptional Field Theory (EFT) applies DFT concepts to exceptional groups  \cite{Berman:2010is,Coimbra:2011nw,Coimbra:2012af,Berman:2012vc,Hohm:2013pua,Hohm:2013vpa,
Hohm:2013uia,Hohm:2014fxa,Berman:2015rcc,Cederwall:2013naa,Cederwall:2015jfa,Arvanitakis_2018,Chabrol:2019kis}. The symmetry of exceptional groups was initially discovered as a partial dimensional symmetry of background fields in 11-dimensional supergravity  \cite{Duff:1985bv,Cremmer:1979up}. 
{
The generalized diffeomorphism in EFT is characterized by the "Y-tensor", which reflects the structure of the exceptional group.
This Y-tensor, $Y^{MN}{}_{PQ}$, is related to the group invariant metric in ${\cal A}$-theory, denoted by $\eta^{MNm}$, through the relation
$Y^{MN}{}_{PQ} = \eta^{MNm} \eta_{PQm}$ where $M,N,\cdots$ are the spacetime indices and $m$ is the  world-volume index. These indices correspond to different representations of the exceptional group.
The origin of the Y-tensor lies in the Schwinger term of the current algebra, $\{\dd_M(\sigma),\dd_N(\sigma')\}= \eta_{MNm}\partial^m
\delta(\sigma-\sigma')$,
where the world-volume derivative $\partial^m$ is defined through the commutator with the Virasoro constraint, ${\cal S}^m= \frac{1}{2}\eta^{MNm}\dd_M\dd_N$.  
The section condition given by the Y-tensor, $Y^{MN}{}_{PQ} \partial_M \partial_N = 0$, is related to the zero-mode of the Virasoro constraint in ${\cal A}$-theory.
Specifically, the zero-mode component of the constraint ${\cal S}^m$ takes the form ${\cal S}^m|_{\rm 0\text{-}modes} = \eta^{MNm} \partial_M \partial_N = 0,$
establishing a direct connection between the section condition in EFT and the Virasoro structure of ${\cal A}$-theory.
}
Active research continues on expressing exceptional groups through brane current algebra \cite{Osten:2021fil,Osten:2023iwc,Osten:2024mjt,Sakatani:2016sko,Blair:2017hhy,Sakatani:2017vbd,Blair:2019tww,Sakatani:2020umt}.

This paper focuses on the 5-brane that describes D=3 ${\cal A}$-theory with SL(5) U-duality symmetry. 
{In the case of D=3, the theory provides a nontrivial yet tractable example 
that includes various types of branes and permits explicit computations. 
The Virasoro algebra, when extended to incorporate brane degrees of freedom, involves Gauss law–type constraints that are intrinsic to the brane. These constraints facilitate the dimensional reduction of the brane world-volume. Such an extended Virasoro algebra serves as a useful prototype for generalization to higher dimensions. Furthermore, the construction of the Lagrangian in this framework constitutes a significant milestone toward extending the formulation to higher-dimensional cases.
}
We clarify how the usual string, and ${\cal T}$-string and membrane of 11-dimensional supergravity (M2) emerge. Specifically, we derive the Lagrangians for the conventional non-perturbative M2-brane, the O(D,D)-covariant ${\cal T}$-string, and the conventional string from the ${\cal A}$-theory 5-brane (${\cal A}$5) Lagrangian.


\subsection{Summary}

In this paper, we focus on the SL(5) U-duality symmetry and clarify the relation between the ${\cal A}$-theory five-brane (${ \cal A}$5-brane) and conventional branes.
The ${\cal A}$5-brane is a 5-brane that exhibits manifest SL(5) U-duality symmetry, and it is described by a perturbative Lagrangian described using the SL(5) rank-two antisymmetric tensor coordinate \cite{Linch:2015fya,Siegel:2020qef,
  Hatsuda:2023dwx}.
The ${\cal A}$5-brane theory is reduced, through the duality reduction procedure, to the ${\cal T}$-string with O(3,3) T-duality symmetry, the ${\cal M}$5-brane with GL(4) duality symmetry, and the $S$tring with GL(3) symmetry. The interrelation among these theories forms a diamond-shaped diagram.
In this paper, we present the sectioning procedure along this diamond contour.
Additionally, we detail reduction procedures from the ${\cal T}$-string Lagrangian to the conventional string Lagrangian, and from the ${\cal M}$5-brane Lagrangian to the conventional non-perturbative M2-brane Lagrangian \cite{Bergshoeff:1987cm}.
We further generalize the dimensional reduction of S-duality, as applied to 11-dimensional supergravity and its relation to type IIA supergravity \cite{Witten:1995ex}, to the case of T-duality, as discussed in subsection \ref{section:1-3}.
For S-duality, $\lambda_{\rm string} \leftrightarrow 1/\lambda_{\rm string}$, the dimensional reduction is realized by taking the limit $\lambda_{\rm string} \ll 1$ in the metric.
For T-duality, characterized by $R/\sqrt{\alpha'} \leftrightarrow \sqrt{\alpha'}/R$ with string length $l_{\rm string} = \sqrt{\alpha'}$, the dimensional reduction is achieved by taking the large-radius limit $R \gg \sqrt{\alpha'}$, such that the O(D,D) spacetime for the ${\cal T}$-string reduces to the conventional D-dimensional spacetime.
Finally, we propose a perturbative ${\cal M}$5-brane Lagrangian in a supergravity background derived from the ${\cal A}$5-brane, as given in equation \bref{M5Lag}.

In section \ref{section:2} 
the relationships among the ${\cal A}$5-brane, ${\cal M}$5-brane, ${\cal T}$-string, and $S$tring theories are elucidated through diamond diagrams based on duality symmetries.
The diamond diagram represents a contour in the U-duality plane, which is parametrized by two quantities: the string coupling and a scale defined by the string length.
The branes are described in terms of field strengths, where the spacetime coordinates possess the gauge symmetry generated by the Gau\ss{} law constraint. 
These field strengths and their associated gauge parameters, along with the world-volume and spacetime coordinates, transform as representations under the relevant duality symmetries.
The world-volume diffeomorphism is generated by the Virasoro constraints ${\cal S}=0$.
Dimensions  to reduce are  determined by
solving the Virasoro constraints  ${\cal S}=0$ 
for spacetime 
and by solving the Gau\ss{} law constraints ${\cal U}=0$ 
for world-volume.

In section \ref{section:3} both the SL(5) and the SL(6) covariant Lagrangians of the ${\cal A}$5-brane are given,
where the SL(6) manifests the 5-brane world-volume Lorentz symmetry.
{
Duff and Lu \cite{DUFF1990394,Duff:2015jka} showed that the membrane theory exhibits the SL(5) duality symmetry by the Gaillard-Zumino approach.
In general, the symmetry of  Lagrangian formulation is larger than that of the corresponding Hamiltonian formulation.
In ${\cal A}$-theory, the U-duality symmetry in the Hamiltonian formulation, G-symmetry, is enhanced to a novel duality symmetry in the Lagrangian formulation,  A-symmetry.
This symmetry enhancement in higher-dimensional cases (D $<$ 6) is summarized on page 6 of \cite{Siegel:2018puf} and page 14 of \cite{Hatsuda:2023dwx}.
It was shown that the brane world-volume metric is also transformed conformally under the SL(5) duality transformation 
as well as the spacetime background fields in \cite{DUFF1990394,Duff:2015jka}.
This mixing between spacetime and world-volume is a manifestation of the extended SL(6) duality symmetry transformation.
}
The SL(6) vielbein includes both the SL(5) spacetime vielbein and the 6-dimensional world-volume vielbein,
so the spacetime and world-volume are mixed under the new duality symmetry SL(6).
The SL(6) formulation is useful to reduce to other branes:
Since the string world-sheet directions and the spacetime directions are direct sum, the SL(6) vielbein is in a block diagonal form
as shown in subsection \ref{section:5-1}.
On the other hand, the brane world-volume directions share the spacetime directions unlike the string
as shown in subsection \ref{section:6-1}.

In section \ref{section:4}, we begin with the O(D,D) string Hamiltonian and apply the double zweibein method \cite{Hatsuda:2018tcx,Hatsuda:2019xiz} to derive the ${\cal T}$-string Lagrangian.
We then present the reduction procedure from the O(D,D) ${\cal T}$-string Lagrangian to the conventional string Lagrangian in D dimensions, following an approach analogous to that in subsection \ref{section:1-3}.
It is shown that the Wess–Zumino term can be obtained by adding a total derivative term.

In section \ref{section:5}, we start from the ${\cal A}$5-brane Lagrangian and present the reduction procedure leading to the ${\cal T}$-string Lagrangian.
The O(D,D) background gauge field is reformulated using SL(4) tensor indices in such a way that it couples naturally to the SL(4) tensor coordinates of the ${\cal T}$-string.
Subsequently, by applying the procedure described in section \ref{section:4}, we derive the conventional string Lagrangian.

In section \ref{section:6}  
we begin with the SL(6) covariant ${\cal A}$5-brane Lagrangian which  leads to a new perturbative ${\cal M}$5-brane Lagrangian.
We further reduce it to the conventional M2-brane Lagrangian. 
{The ``perturbative" ${\cal M}$5-brane Lagrangian is formulated as a bilinear expression in terms of currents, while the ``non-perturbative" M2-brane Lagrangian comprises the sum of the Nambu–Goto and Wess–Zumino terms.
The dimensional reduction from the ${\cal M}$5-brane to the M2-brane is implemented via 
the ``non-perturbative projection" $ \partial^m = \epsilon^{ij} \partial_j x^m \partial_i, $
 in \bref{stwvMix} and  the gauge fixing of the world-volume metric in \bref{gaugechoicewv}. 
}
The Nambu-Goto Lagrangian is obtained by the gauge choice of the world-volume vielbein,
while the Wess-Zumino term is obtained by adding the total derivative term.

	\subsection{Dimensional reduction procedure}\label{section:1-3}

In \cite{Witten:1995ex}, it was pointed out that under an S-duality transformation between the 10-dimensional type IIA theory and the 11-dimensional supergravity theory, the structure of the supersymmetry algebra remains invariant, although the interpretation of the central charge changes. The global superalgebra, involving supercharges $Q$ and $Q'$ of the opposite chirality, 10-dimensional momenta $P$, and a central charge $W$, is given by
$\{Q, Q\} \sim P \sim \{Q', Q'\}, \quad \{Q_\alpha, Q'_{\dot{\beta}}\} \sim \delta_{\alpha\dot{\beta}} W$.
The central charge $W$ is interpreted as the Ramond-Ramond (RR) D0-brane charge in 10 dimensions, and as the momentum in the 11th dimension in 11-dimensional supergravity.
The 11-dimensional spacetime reduces into the 10-dimensional spacetime
 in the weak coupling limit 
 $e^{2\phi}\ll 1$,
\bea
ds^2_{11}=g^{[10]}_{mn}dx^mdx^n+e^{2\phi}(dy-A_mdx^m)^2~\xrightarrow[\rm reduction]{\rm dimensional}~
ds^2_{10}=g^{[10]}_{mn}dx^mdx^n
\eea
with the 11-th dimensional coordinate $y$ and the string coupling   $\lambda_{\rm string}=e^{3\phi/2}$.
The 11-dimensional momentum $W$ is maintained as the D0-brane charge in the 10-dimeniosnal IIA theory after the dimensional reduction.

This framework is generalized to incorporate T-duality.
We compare the superalgebra of the $2D$-dimensional ${\cal T}$-string theory, which exhibits manifest T-duality, with the type II superalgebra of conventional string theory in $D$ dimensions.
The global type II superalgebra involves two supercharges, $Q$ and $Q'$, and the $D$-dimensional momentum $P$, and is expressed as
$\{Q, Q\} \sim (P + \tilde{P}), \quad \{Q', Q'\} \sim (P - \tilde{P})$, 
where $\tilde{P}$ is a central charge.
In $D$ dimensions, $\tilde{P}$ is interpreted as an NS–NS charge, while in $2D$ dimensions, it corresponds to the additional momenta associated with the extended spacetime.
By restoring the $\alpha'$ dependence in the O($D,D$) momentum–winding vector,
$(p_m,~\frac{1}{\alpha'} \partial_\sigma x^m) \rightarrow (p_m,~\frac{1}{\alpha'} \tilde{p}^m)$, 
the canonical conjugate coordinates become $(x^m,~\alpha' y_m)$.
At small compactification scales ($R \ll \sqrt{\alpha'}$), the winding modes become light and are readily excited, whereas at large scales ($R \gg \sqrt{\alpha'}$), they become heavy and only the momentum modes remain dynamically relevant.
Manifest T-duality is broken by choosing a specific background such that the $2D$-dimensional spacetime effectively reduces to the $D$-dimensional one in the limit $R \gg \sqrt{\alpha'}$:
\bea
&ds^2_{\rm 2D}=g
_{mn}dx^mdx^n+\alpha'{}^2(dy_m-dx^lB_{lm})g
^{mn}(dy_n-dx^kB_{kn})&\nn\\
&~~~~~~~~~~~~~~~~~~~~~~~~\xrightarrow[\rm reduction]{\rm dimensional}~
ds^2_{\rm D}=g
_{mn}dx^mdx^n \label{dimreduction}~~~.&
\eea
Here, $y_m$ denotes the additional $D$-dimensional coordinates, and $B_{mn}$ is the NS–NS gauge field.
The additional momenta in the extended dimensions are preserved as NS–NS charges after the dimensional reduction.

{This dimensional reduction procedure corresponds to the gauge fixing of the dimensional reduction constraint which is the first class constraint
in Hamiltonian formulation. 
The dimensional reduction constraint 
and the gauge fixing to reduce the conventional string 
are discussed in \cite{Hatsuda_2015} for a flat space case.
The dimensional reduction constraint is the $y$ component of the symmetry generator, $\tilde{\dd}_y=0$. The gauge fixing condition $\partial_\sigma y=0$ reduces the set of conventional string operators,
the physical momentum ${P}_x\neq 0$ and left/right covariant derivatives
$P_x\pm \partial_\sigma x$.
In Lagrangian formulation the momentum is replaced by  $P_{X}=\partial L/\partial \dot{X}$.
It is generalized to the brane case, and then the dimensional reduction constraint
turns out to be the Virasoro constraint in which one of the momenta is replaced by the 0-mode \cite{Linch:2015qva}. For zero-mode momenta $p_x,~p_y$ 
and momenta including both the 0-mode and the non-0-modes  $P_x,~P_y$ , 
 the dimensional reduction constraint $\tilde{\dd}_y=0$ is expressed by the Virasoro operator as
$p_{y}\cdot P_{x}+ p_{x}\cdot P_{y}=0 $ $\to$ $P_{y}=0$.}

{Although the equation of motion derived from the doubled Lagrangian, when combined with the self-duality condition, coincides with that obtained from the original Lagrangian, the self-duality condition causes the doubled Lagrangian to vanish \cite{Cremmer_1998}.
In particular, the self-duality condition $\partial_\mu x = \epsilon_{\mu\nu} \partial^\nu y$ reduces the Lagrangian of the O($D,D$) ${\cal T}$-string in flat space to zero, as follows
\bea
L&=&\frac{1}{2}\left(\dot{x}^2-x'^2+\dot{y}^2-y'^2\right)~
\xrightarrow{\rm selfduality}~0\nn~~~.\label{flatTst}
\eea
 It is also mentioned that the naive section $y=0$ the ${\cal T}$-string Lagrangian in curved background 
does not reduce to the expected string Lagrangian in curved background as
\bea
L&=&\frac{1}{2}\left(
\partial_+ x^m~~\partial_+ y_m
\right)
\left(\begin{array}{cc}g_{mn}-B_{ml}g^{lk}B_{kn}
	&-B_{ml}g^{ln}\\g^{ml}B_{ln}&g^{mn}\end{array}\right)
\left(\begin{array}{c}
	\partial_- x^n\\\partial_- y_n\end{array}\right)\nn\\
&&~
\xrightarrow{y=0}~\frac{1}{2}
\partial_+ x^m 
( g_{mn}-B_{ml}g^{lk}B_{kn})	\partial_- x^n
\nn~~~.
\eea

The following points are also noteworthy.
Integrating out the $(dy + \cdots)^2$ term is possible in the case of a constant background. However, for a general non-constant background, the path integral over $dy$ in
$\exp\left[-\int (dy + \cdots)\, g\, (dy + \cdots)\right]$
yields a Jacobian factor $\sqrt{g}$ in the path integral measure, which in turn generates an additional term in the effective action.
Using the equation of motion is again valid in the constant background case, but it does not reproduce the conventional string Lagrangian when the background fields $g(x,y)$ and $B(x,y)$ are non-constant.
Furthermore, imposing both conditions
$\partial_+ y - \partial_+ x\, B = 0 \quad \text{and} \quad \partial_- y - \partial_- x\, B = 0$
is inconsistent, since the integrability condition is violated in curved backgrounds where $[\partial_+, \partial_-] y \neq 0$.
Several studies have been devoted to refining the reduction to the conventional string Lagrangian, resulting in a variety of interesting approaches
\cite{Cremmer_1998, Hull_2005, Hull_2007, Lee_2014, Sakatani:2016sko, Arvanitakis_2018}.

Instead we propose the reduction procedure from ${\cal T}$-string Lagrangian to the conventional string Lagrangian:
(1) adding the total derivative term
${-}\partial_\mu(\epsilon^{\mu\nu}x^m\partial_\nu y_m)$ 
to derive the Wess-Zumino term, then 
(2) the dimensional reduction \bref{dimreduction}
as
\bea
L&=&\frac{1}{2}\left(
	\partial_+ x^m~~\partial_+ y_m
\right)
\left(\begin{array}{cc}g_{mn}-B_{ml}g^{lk}B_{kn}
	&-B_{ml}g^{ln}\\g^{ml}B_{ln}&g^{mn}\end{array}\right)
\left(\begin{array}{c}
	\partial_- x^n\\\partial_- y_n\end{array}\right)\nn
{-}\partial_\mu(\epsilon^{\mu\nu}x^m\partial_\nu y_m)\nn\\
&&~\xrightarrow[\rm reduction~(1.2)]{\rm dimensional}~
\partial_+ x^m(g_{mn}+B_{mn})	\partial_- x^n
\label{dimredWZ}~~~.
\eea
This is the expected string  Lagrangian up to the normalization factor two which can be absorbed by the Lagrange multiplier.
The section conditions of spacetime fields $\Phi(x,y)$ are consistent with the Lagrangian where the section $y=0$ can be chosen as $\Phi(x)$. 

This procedure is similar to the usual dimensional reduction where the reduction is done in the local flat Lorentz coordinate. i.e. Suppose that we have a line element $dx^A \equiv  dx^{M} E_M{}^{A}$. We decompose 
{the doubled coordinate $dx^A$ into $dx^a$ and $dy_a$
	in the local Lorenz frame, and then discard $(dy_a)^2$.}
Since the metric ($\hat{\eta}$-tensor) in local flat spacetime is already diagonal, in practice we can just apply this reduction by deleting certain blocks of $\hat{\eta}$-tensor similar to \bref{dimreduction}.

\par\vskip 6mm
The main purpose of this paper is to carry out the above reduction procedure in several specific cases. In general, however, the procedure can be schematically summarized as follows.

\begin{enumerate}
\item  We start with the current algebra 
defined on an extended space of coordinates, where both momentum and winding modes have their corresponding conjugate coordinates. 
The Hamiltonian is written as a sum of self-dual and anti-self-dual constraints:
 $H = g{\cal H} + \tilde{g}\tilde{\mathcal{H}} + s_{m}{\cal S}^{m} + \tilde{s}_{m}\tilde{{\cal S}}^{m} + Y^{m}{\cal U}_{m}$, where ${\cal H}, \tilde{{\cal H}}$ are the $\tau$ component of the Virasoro constraints and its dual counterpart, ${\cal S}^{m}, \tilde{{\cal S}}^{m}$ are the $\sigma^m$ components of the Virasoro constraints and its dual counterpart.  ${\cal U}_{m}$ is  the Gau\ss{} law constraint specific to branes.
\item 
The Lagrangian is obtained via a Legendre transformation of the Hamiltonian $H$. This has been performed in previous works for various theories \cite{Hatsuda:2023dwx}. Schematically, the Lagrangian takes the form
 $L =\Phi J_{\textrm{SD}}\cdot \hat{\eta}\cdot J_{\overline{\textrm{SD}}} + \Lambda \cdot J_{\overline{\textrm{SD}}}\cdot \hat{\eta}\cdot J_{\overline{\textrm{SD}}}+\cdots$, where $J^{A}_{\textrm{SD}/\overline{\textrm{SD}}}$ are the selfdual and anti-selfdual currents. They are coupled with vielbein, and thus they have flat indices. $\Phi$ and $\Lambda$ are Lagrange multipliers which are  functions of $g, s^{m}, \tilde{g}, \tilde{s}^{m}$. One can gauge fix $\Lambda=0$ by the suitable choice of original parameters.

\item 
Separate coordinates and  currents into the physical part and the auxiliary part
as $X^{M} \rightarrow x^{m}, y^{\mu}$, and $J^{A}\rightarrow J^{a}, J^{\alpha}$ 
where $x^m$ represents the physical coordinates for the target string or brane theory, and $y^\mu$ denotes auxiliary coordinates. Dimensional reduction is then performed according to equation \eqref{dimreduction}.

\item 
The reduced Lagrangian $L{}' = J^{a}{}_{\textrm{SD}}\hat{\eta}_{ab}J^{b}{}_{\overline{\textrm{SD}}}$ 
can be shown to reproduce the desired string or brane action, up to the absence of the Wess–Zumino (WZ) term.  
We find that adding a total derivative term to the Lagrangian restores the WZ term
	$$
	L{} + \textrm{Total derivative}  =  J^{a}{}_{\textrm{SD}}\hat{\eta}_{ab}J^{b}{}_{\overline{\textrm{SD}}} + \tilde{J}^{\alpha}_{\textrm{SD}}\hat{\eta}_{\alpha\beta}\tilde{J}^{\beta}{}_{\overline{\textrm{SD}}} + L_{\textrm{WZ}}.
	$$
Here, the current $\tilde{J}^\alpha$ is modified by the addition of the total derivative term and is subsequently eliminated through the dimensional reduction \bref{dimreduction}. This procedure yields the correct WZ term.
\end{enumerate}

\section{Theories with manifest duality symmetries and sectionings}\label{section:2}
\par
\subsection{Diamond diagrams}
The duality web of the $G$-symmetry in ${\cal A}$-theory is represented by the diamond diagram shown in Fig. \hyperref[DAGATMSWeb]{1}, as studied in \cite{Linch:2015qva}.
The $G$-symmetry, associated with the coset group $G/H$, plays the role of a duality symmetry.
The coset parameter serves as the gauge field of the duality-covariant geometry, incorporating the spacetime vielbein as well as the NS–NS and R–R gauge fields of superstring theory.

The relationships among these duality groups are illustrated using Dynkin diagrams, as discussed in \cite{Hatsuda:2023dwx}.
Removing a single node from the Dynkin diagram of E$_{\rm{D+1(D+1)}}$ reduces it to that of either {GL}(D+1) or O(D,D), depending on which node is removed.
Further removing one more node from the Dynkin diagram of {GL}(D+1) or O(D,D) leads to that of {GL}(D).

\bea
&\begin{array}{c}
	\begin{array}{c}
		\cal{A}\mathchar`-\rm{theory}\\		\rm{E}_{{D}+1({D}+1)}/\rm{H}_{{D}}\\
		\rm{bispinor}
	\end{array}
	\\
	~	\swarrow\quad\quad\quad\quad\quad\quad\quad\quad
	\searrow~
	\\
	\begin{array}{c}
		\cal{M}\mathchar`-\rm{theory}\\
		\rm{GL}(D+1)/\rm{SO}(D)\\
		{D}+1
	\end{array}
	\quad\quad\quad
	\begin{array}{c}
		\cal{T}\mathchar`-\rm{theory}\\
		\rm{O}(D,D)/\rm{SO}(D)^2\\
		2{D}
	\end{array}\\
~	\searrow\quad\quad\quad\quad\quad\quad\quad\quad\swarrow~
\\
	\begin{array}{c}
		\rm{S}\mathchar`-\rm{theory}\\
		\rm{GL}(D)/\rm{SO}(D)\\
		{D}
	\end{array}
\end{array}\nn&\\\label{DAGATMSWeb}
\nn
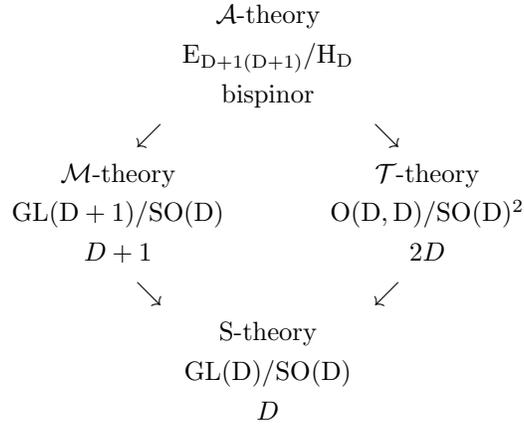
\\&{\rm{Figure~1:}} ~G\mathchar`-{\rm symmetries~ of}~ {\rm{D}} ={D}~\rm{theories~and~spacetime~dimensions}& \nn
\eea

In this paper we focus on D=3 case where the $G$-symmetry is SL(5) and the diamond diagram becomes Fig. \hyperref[Fig2]{2}.
This SL(5) duality symmetry is enlarged to SL(6) for the (5+1)-dimensional world-volume covariance in Lagrangian \cite{Hatsuda:2023dwx}. 
We named this enlarged symmetry ``$A$-symmetry".
This ${\cal A}$-theory unifies the spacetime and the world-volume, 
in a sense that the coset parameter of $A/L$=SL(6)/GL(4) includes not only  the spacetime vielbein field but also the world-volume vielbein field.

In this paper, we focus on the D=3 case, where the $G$-symmetry is SL(5) and the diamond diagram corresponds to Fig. \hyperref[Fig2]{2}.
This SL(5) duality symmetry is further enhanced to SL(6) in order to accommodate the $(5+1)$-dimensional world-volume covariance in the Lagrangian formulation \cite{Hatsuda:2023dwx}.
We refer to this enlarged symmetry as the ``$A$-symmetry".
The resulting ${\cal A}$-theory unifies the spacetime and world-volume structures, in the sense that the coset parameter of $A/L = ${SL}(6)/{GL}(4) includes not only the spacetime vielbein field but also the world-volume vielbein field.

\bea
&\begin{array}{c}
	\begin{array}{c}
		\cal{A}\mathchar`-\rm{theory}\\		
		{ A/L}={	\rm{SL}(6)}/{\rm{GL}(4)}
	\\15\\
		{ G/H}=
		{	\rm{SL}(5)}/{\rm{SO}(5)}\\
		10
	\end{array}
	\\
	~	\swarrow\quad\quad\quad\quad\quad\quad\quad\quad
	\searrow~
	\\
	\begin{array}{c}
		\cal{M}\mathchar`-\rm{theory}\\
		{ G/H}={	\rm{GL}(4)}/{\rm{SO}(4)}\\4
	\end{array}
	\quad\quad\quad
	\begin{array}{c}
		\cal{T}\mathchar`-\rm{theory}\\
		{ G/H}={		\rm{O}(3,3)}/{\rm{SO}(3)^2}\\3+3
	\end{array}\\
	~	\searrow\quad\quad\quad\quad\quad\quad\quad\quad\swarrow~
	\\
	\begin{array}{c}
		\rm{S}\mathchar`-\rm{theory}\\
		{G/H}={		\rm{GL}(3)}/{\rm{SO}(3)}\\3
	\end{array}
\end{array}&\nn\\\label{Fig2}\nn
\\&{\rm{Figure}}~2: {A}\mathchar`-~{\rm{and}} ~G\mathchar`-{\rm symmetries ~of~D=3}~{\rm theories~and~spacetime~dimensions}&\nn
\eea
We note that $H =$ {SO}(D) is used instead of {SO}(D-1,1) for simplicity. Consequently, a Wick rotation is required to properly account for the time component in this section and elsewhere.
\par
\subsection{Representations}

In duality covariant theories, spacetime and world-volume coordinates transform as representations of the duality symmetry ($A$-symmetry or $G$-symmetry), which determines the world-volume dimension. The Gau\ss{} law constraint generates gauge symmetry of the duality covariant spacetime coordinate, making the brane current correspond to a field strength.

The D=3 ${\cal M,~T},~S$-theories are obtained from
the D=3 ${\cal A}$-theory \cite{Linch:2015fya, Siegel:2020qef}.
We list representations of duality groups in Table \hyperref[rep]{1};
the world-volume derivative $\partial^{m}$,
the gauge parameter $\lambda^{m}$,
the spacetime coordinate $X^M$,
and the field strength (the current) $F_M=\eta_{MNm}\partial^m X^{N}$ ( $J_\mu{}^M=\partial_\mu X^M$, $\mu=(\tau, \sigma)$).
$\eta_{MNm}$ is the $G$-symmetry invariant tensor which enters the current algebra,
in which the SL(5) invariant metric is $\eta_{MNl}=\eta_{m_1m_2n_1n_2l}=\epsilon_{m_1m_2n_1n_2l}$.
\bea 
&{\renewcommand{\arraystretch}{1.5}
	\begin{array}{|c|c|c|c|c|c|}
		\hline
		\begin{array}{c}
			{\rm Theories}\\{\rm Groups}\end{array}&	
		\begin{array}{c}{\rm World}\mathchar`-{\rm volumes}\\\partial\end{array}
		&\begin{array}{c}{\rm Gauges}\\\lambda\end{array}&
		\begin{array}{c}{\rm Spacetimes}\\X\end{array}&
		\begin{array}{c}{\rm Field~strengths}\\F\end{array}\\\hline\hline
		{\cal A}\mathchar`-{\rm theory}&		6&6&15&20\\
		{\rm SL}(6)&		\partial^{\hat{m}}, ~_{\hat{m}=0,\cdots,5}&\lambda^{\hat{m}}  &X^{\hat{m}\hat{n}}&F^{\hat{m}\hat{n}\hat{p}} \\		\hline	
		{\cal A}\mathchar`-{\rm theory}&	1\oplus 5&1\oplus 5&10\oplus 5&10\oplus 10'\\
		{\rm SL}(5)&\partial^{0}, \partial^{m},~_{m=1,\cdots,5}& \lambda^0, \lambda^m  &X^{mn},Y^m&F_\tau{}^{mn}, F_\sigma{}_{mn}\\
		\hline	
		{\cal M}\mathchar`-{\rm theory}&	1\oplus 4(5)&1\oplus 4&
		4\oplus 1
		&4\oplus 6
		\\
		{\rm GL}(4)&\partial^{0}, \partial^{\underline{m}},~_{\underline{m}=1,\cdots,4}& \lambda^0, \lambda^{\underline{m}}  &x^{\underline{m}},Y
		&F_\tau{}^{\underline{m}},
		F_\sigma{}_{\underline{m}\underline{n}}
		\\
		\hline	
		{\cal T}\mathchar`-{\rm theory}&	1\oplus 1&0&3\oplus 3'&3\oplus 3\oplus 3'\oplus 3'\\
		{\rm O}(3,3)&\partial^{0}, \partial^{\sigma},~_{\bar{m}=1,2,3}&   &x^{\bar{m}},y^{\bar{m}\bar{n}}&J_\tau{}^{\bar{m}},J_\sigma{}_{\bar{m}\bar{n}};J_\tau{}^{\bar{m}},J_\sigma{}_{\bar{m}},\\
		\hline	
		S\mathchar`-{\rm theory}&	1\oplus 1&0&3&3\oplus 3\\
		{\rm GL}(3)&\partial^{0}, \partial^{\sigma},~_{\bar{m}=1,2,3}&   &x^{\bar{m}}&J_\tau{}^{\bar{m}},J_\sigma{}_{\bar{m}\bar{n}}\\
		\hline	
	\end{array}\nn}&\\\label{rep}\nn\\&{\rm{Table}}~1: {\rm Representations~of~the~duality~groups~of~theories}&\nn
\eea
The  world-volume  dimension of ${\cal M}$-theory is still $1\oplus 5$ where four dimensions are embedded in the 4 spacetime $x^{\underline{m}}$
and one dimension is embedded in the internal space,
so we denote as $1\oplus 4(5)$.

The field strengths and currents together with the gauge transformations are given concretely  as follows.
\begin{enumerate}
\item{${\cal A}$5-brane field strengths
	
\begin{enumerate}
	\item {World-volume covariant ${\cal A}$5-brane field strength 
		
		The SL(6) $A$-symmetry covariant ${\cal A}$-theory is described by a 5-brane with the manifest SL(6) new duality symmetry which manifests 6-dimensional world-volume Lorentz symmetry, namely world-volume covariant ${\cal A}$5-brane.
		\bea
		&
		F^{\hat{m}\hat{n}\hat{p}}=\frac{1}{2}\partial^{[\hat{m}}X^{\hat{n}\hat{p}]}~~,~~
		\delta_\lambda X^{\hat{m}\hat{n}}=\partial^{[\hat{m}}\lambda^{\hat{n}]}~~,~~
		\hat{m}=0,1,\cdots,5
		\label{SL6F}&
		\eea
	}
	\item{${\cal A}$5-brane field strength 
		
		The SL(5) $G$-symmetry covariant  ${\cal A}$-theory is described by a 5-brane with manifest SL(5) U-duality symmetry, namely ${\cal A}$5-brane.
		\bea
		{\renewcommand{\arraystretch}{1.8}
			\left\{\begin{array}{l}
				F_\tau{}^{mn}=\dot{X}^{mn}-\partial^{[m}Y^{n]}\\
				F_\sigma{}_{;m_1m_2}=\frac{1}{2}\epsilon_{m_1\cdots m_5}\partial^{m_3}X^{m_4m_5}\\
				m=1,\cdots,5\end{array}\right. , ~
			\left\{\begin{array}{l}
				\delta_\lambda X^{mn}=\partial^{[m}\lambda^{n]}\\
				\delta_\lambda Y^{m}=\dot{\lambda}^m-\partial^{m}\lambda^0
			\end{array}\right.}\label{SL(5)A5current}
		\eea}
\end{enumerate}
	}
	\item{${\cal M}$5-brane field strength

		The GL(4) ${\cal M}$-theory is described by a 5-brane with the manifest GL(4) duality symmetry, namely ${\cal M}$5-brane.
		We focus only on 4-dimensional subspace of the 5-dimensional world-volume which is embedded in the 4-dimensional spacetime. 
{       This ${\cal M}$5-brane extends over both the main space (i.e., the duality-covariant space) and the internal space.
Four of its world-volume directions lie in the main space while the remaining directions lie in the internal space, specifically one world-volume direction in the Hamiltonian formalism, or two in the Lagrangian formalism.
Considering the critical string action in the full spacetime structure is an interesting subject, although it lies beyond the scope of the present discussion.}
{The relationship between the main space and the internal space is schematically illustrated in Figure 2 the  ``slug diagram" (see page 27 in \cite{Linch:2016ipx} or page 14 in \cite{Hatsuda:2023dwx}).
In the case of D = 3 the main space coordinate is represented by a bispinor $X^{\alpha\beta}$, and the world-volume coordinate by an antisymmetric bispinor $\sigma^{[\alpha\beta]}$, with $\alpha = 1, \dots, 4$.
The internal space coordinate is given by a bispinor $Y^{[\alpha'\beta']}$, where $\alpha' = 1, \dots, 8$.
The total number of supersymmetries is 32, which corresponds to the product of the dimensions of the spinor indices $32 = 4 \times 8$.}
{It is noted that the assignment of the duality symmetric space in ${\cal A}$-theory differs from that in conventional formulations.
In ${\cal A}$-theory, the duality-symmetric space is assigned to the main "spacetime" rather than the internal space, such that all tensor gauge fields are automatically incorporated into the coset parameter of $\mathrm{E}_{\mathrm{D+1}} / H$.        }
\\ \\
		Physical currents are as follows.
		\bea
		{\renewcommand{\arraystretch}{1.8}
			\left\{\begin{array}{l}
				F_\tau{}^{\underline{m}}=\dot{x}^{\underline{m}}+\partial^{\underline{m}}Y\\
				F_\sigma{}_{;\underline{m}_1\underline{m}_2}
				=-\epsilon_{\underline{m}_1\cdots \underline{m}_4}\partial^{\underline{m}_3}x^{\underline{m}_4}\\
				\underline{m}=1,\cdots,4
			\end{array}\right. , ~
			\left\{\begin{array}{l}
				\delta_\lambda x^{\underline{m}}=\partial^{\underline{m}}\lambda\\
				\delta_\lambda Y=-\dot{\lambda}
			\end{array}\right.}\label{GL(4)M52current}
		\eea
		The following currents are auxiliary written by auxiliary coordinates $y^{\underline{mn}}$, $Y^{\underline{m}}$. 
		\bea
		{\renewcommand{\arraystretch}{1.8}
			\left\{\begin{array}{l}
				F_\tau{}^{\underline{m}\underline{n}}
				=\dot{y}^{\underline{m}\underline{n}}
				-\partial^{[\underline{m}}Y^{\underline{n}]}\\
				F_\sigma{}_{;\underline{m}_1}
				=\frac{1}{2}\epsilon_{\underline{m}_1\cdots \underline{m}_4}\partial^{\underline{m}_2}y^{\underline{m}_3\underline{m}_4}
			\end{array}\right. , ~
			\left\{\begin{array}{l}
				\delta_\lambda y^{\underline{m}\underline{n}}=\partial^{[\underline{m}}\lambda^{\underline{n}]}\\
				\delta_\lambda Y^{\underline{m}}=\dot{\lambda}^{\underline{m}}\\	\delta_\lambda \lambda^{\underline{m}}=\partial^{\underline{m}}{\lambda}
			\end{array}\right.}
		\eea
		These currents constitute the SL(5) $A$-symmetry together with \bref{GL(4)M52current}, and they are used to lead the non-perturbative M2-brane Lagrangian.
}
	\item{${\cal T}$-string currents
		
		The O(3,3) ${\cal T}$-theory is described by a string with the manifest O(3,3) T-duality symmetry, namely ${\cal T}$-string.
		\bea
	{\renewcommand{\arraystretch}{1.8}
		\left\{\begin{array}{l}
			J_\tau{}^{\underline{m}_1\underline{m}_2}
			=\dot{X}^{\underline{m}_1\underline{m}_2}\\
			J_\sigma{}_{\underline{m}_1\underline{m}_2}
			=\frac{1}{2}\epsilon_{\underline{m}_1\cdots \underline{m}_4}\partial_{\sigma}X^{\underline{m}_3\underline{m}_4}
		\end{array}\right. }
	\eea
It is convenient to represent in terms of $x^{\bar{m}}$ and $y_{\bar{m}}=\frac{1}{2}\epsilon_{\bar{m}\bar{n}\bar{l}}y^{\bar{n}\bar{l}}$.
		\bea
		{\renewcommand{\arraystretch}{1.8}
			\left\{\begin{array}{l}
				J_\tau{}^{\bar{m}}=\dot{x}^{\bar{m}}\\
				J_\sigma{}_{\bar{m}_1\bar{m}_2}
				=-\epsilon_{\bar{m}_1\bar{m}_2 \bar{m}_3}\partial_{\sigma}x^{\bar{m}_3}\\
				J_\tau{}^{\bar{m}_1\bar{m}_2}
				=\dot{y}^{\bar{m}_1\bar{m}_2}\\
				J_\sigma{}_{\bar{m}_1}
				=\frac{1}{2}\epsilon_{\bar{m}_1\bar{m}_2 \bar{m}_3}\partial_\sigma y^{\bar{m}_2\bar{m}_3}\\
				\bar{m}=1,2,3
			\end{array}\right. }
		\eea	}
	\item{$S$-tring currents

		The GL(3) $S$-theory is described by a string with the manifest GL(3)  spacetime diffeomorphism symmetry, namely a 3-dimensional string.
		\bea
		{\renewcommand{\arraystretch}{1.8}
			\left\{\begin{array}{l}
				J_\tau{}^{\bar{m}}=\dot{x}^{\bar{m}}\\
				J_\sigma{}_{\bar{m}_1\bar{m}_2}
				=-\epsilon_{\bar{m}_1\bar{m}_2 \bar{m}_3}\partial_{\sigma}x^{\bar{m}_3}
			\end{array}\right. }
		\eea
	}
\end{enumerate}
Some minus signs come from the mere notation $\epsilon_{1234}=1=-\epsilon_{4123}$.
It is denoted that these currents are flat currents,
and in later sections flat current symbols $\don\circ{F}$ or $\don\circ{J}$ will be used to distinguish from curved background currents.

\par
\vskip 6mm
\par
\vskip 6mm
\subsection{Constraints and sectionings}

The theories in Hamiltonian formulation are constructed by the current algebra with manifest duality symmetries \cite{Linch:2015fya}.
The Spacetime translation is generated by the covariant derivative $\dd_M(\sigma)$.
The  $p$-brane current algebra with $G$-symmetry covariance is given by 
\bea
\left[\dd_{M}(\sigma),\dd_{N}(\sigma')\right]
&=&2if_{MN}{}^L\dd_L(\sigma)\delta(\sigma-\sigma')+
2i\eta_{MNm}\partial^{m}\delta^{(p)}(\sigma-\sigma')
		\label{GCA}~~~.
\eea
Branes are governed by the brane Virasoro constraints ${\cal S}^m=\frac{1}{2}\dd_M\eta^{MNm}\dd_N=0$
and ${\cal H}=\frac{1}{2}\dd_M\hat{\eta}^{MN}\dd_N=0$ together with the Gau\ss{} law constraints ${\cal U}_m=0$
which is required by the  closure of 
the Virasoro algebra.
$\hat{\eta}^{MN}$ is the $H$-invariant metric.
Theories are related by sectionings; 
The Virasoro constraint  ${\cal S}^m=0$ gives the section conditions to reduce the spacetime dimensions, and the Gau\ss{} law constraint ${\cal U}_m=0$ is  used to reduce the world-volume dimension as Fig. \hyperref[AGATMSWebsec]{3}.
\bea
&\begin{array}{c}
	\begin{array}{c}
		{\cal A}5\mathchar`-\rm{brane}\\
		10\mathchar`-\rm{dim.~spacetime}
	\end{array}
	\\\\
	{\cal S}~	\swarrow\quad\quad\quad\quad\quad\quad\quad\quad
	\searrow~{\cal U}\\\\
	\begin{array}{c}
		{\cal M}5\mathchar`-\rm{brane}\\4\mathchar`-\rm{dim.~spacetime}
	\end{array}
	\quad\quad\quad
	\begin{array}{c}
		{\cal T}\mathchar`-\rm{string}\\	3+3\mathchar`-\rm{dim.~spacetime}
	\end{array}\\\\
	{\cal U}~	\searrow\quad\quad\quad\quad\quad\quad\quad\quad\swarrow~{\cal S}\\\\
	\begin{array}{c}
		S\rm{tring}\\	3\mathchar`-\rm{dim.~spacetime}
	\end{array}
\end{array}\label{AGATMSWebsec}&\nn\\\nn
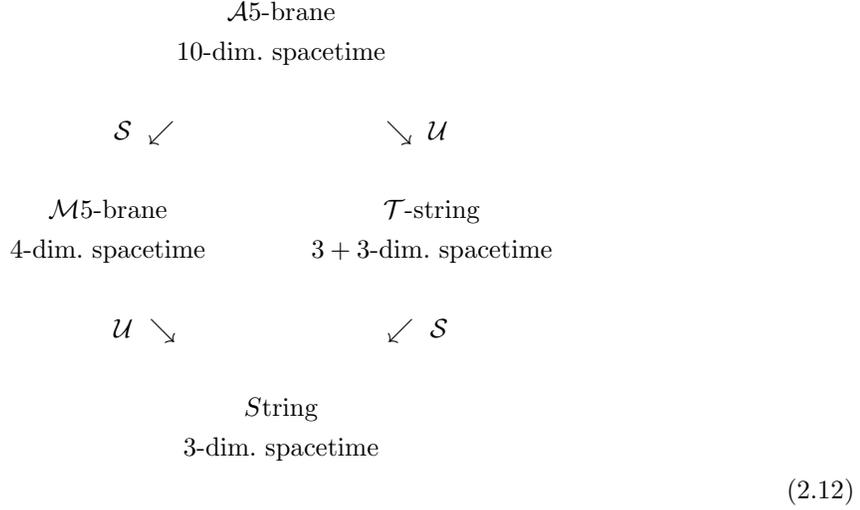
\\&\rm{Figure}~3: {\rm Diamond~diagram~of~Sectionings~of} ~{branes ~of ~D=3}~{\rm theories}&\nn
\eea

The spacetime covariant derivatives, constraints and section conditions are given \cite{Linch:2015fya, Siegel:2020qef} in Fig.\hyperref[AGATMSWebsec]{3} concretely as follows.
\begin{enumerate}
	\item {${\cal A}$5-brane in 10-dimensional spacetime

		The 10-dimensional spacetime is described by the 
        rank-two anti-symmetric tensor covariant derivative $\dd_{m_1m_2}$ as
		\bea
		\dd_{m_1m_2}&=&P_{m_1m_2}+\frac{1}{2}\epsilon_{m_1\cdots m_5}
		\partial^{m_3}X^{m_4m_5}\label{SL5ACA}
		\eea
		where  $P_{mn}$ is canonical conjugate of $X^{mn}$ with
		$[P_{mn}(\sigma),X^{lk}(\sigma')]=\frac{1}{i}\delta_{m}^{[l}\delta_n^{k]}\delta^{(5)}(\sigma-\sigma')$ and $m=1,\cdots,5$.
		
		The SL(5) covariant current algebra of ${\cal A}$5-brane is given by
		\bea
		\left[\dd_{{m}_1{m}_2}(\sigma),\dd_{{m}_3 {m}_4}(\sigma')\right]
		&=&2i\epsilon_{{m}_1\cdots {m}_5}
		\partial^{m_5}\delta^{(5)}(\sigma-\sigma')
		\label{SL5A5CA}~~~.
		\eea
		
The 5-dimensional world-volume diffeomorphism is generated by the Virasoro constraints ${\cal S}^m=0$ while the world-volume time diffeomorphism is generated by ${\cal H}=0$. The Gau\ss{} law constraint ${\cal U}_m=0$ generates the gauge symmetry of the spacetime coordinate. These constraints are given by \cite{Hatsuda:2023dwx} as:
		\bea
		{\renewcommand{\arraystretch}{1.8}
			\left\{
			\begin{array}{ccl}
				{\cal S}^m&=&\frac{1}{16}\dd_{m_1m_2}\epsilon^{mm_1\cdots m_4}\dd_{m_3m_4}=0\\
				{\cal H}&=&\frac{1}{16}\dd_{m_1m_2}\delta^{m_1[n_1}\delta^{n_2]m_2}\dd_{n_1n_2}=0\\
				{\cal U}_m&=&\partial^n\dd_{mn}=0
			\end{array}
			\right.\label{VirasoroSL5}}
		\eea

The SL(5) covariant constraints ${\cal S}^m=0$ and ${\cal U}_m=0$ are background independent. 
These constraints are used as the dimensional reduction and section condition by replacing the spacetime momentum $P_M(\sigma)$ 
with the derivative of the 0-mode of the spacetime coordinate 
$X_{ 0}^M$. 
\bea
&{\renewcommand{\arraystretch}{3.0}
\begin{array}{|l|c|c|}
	\hline
&{\rm Virasoro}:~ {\cal S}^m&{\rm Gau\ss{}~law}:~{\cal U}_m\\
\hline
\begin{array}{l}
{\rm dimensional}~
{\rm reduction}
\end{array}
&	\displaystyle\epsilon^{mm_1\cdots m_4}P_{m_1m_2}(\sigma)\frac{\partial}{\partial X^{m_3m_4}_0}
	&P_{mn}(\sigma)\displaystyle\frac{\partial}{\partial \sigma_n}\\\hline
\begin{array}{l}
	{\rm section}~
	{\rm condition}
\end{array}&
\displaystyle\epsilon^{mm_1\cdots m_4}\frac{\partial}{\partial X^{m_1m_2}_0}\frac{\partial}{\partial X^{m_3m_4}_0}
&\displaystyle\frac{\partial}{\partial \sigma_n}
\frac{\partial}{\partial X^{mn}_0}\\\hline
\end{array}}
\label{SectionVira}\label{sectionGau}&\nn\\\nn\\
&{\rm Table}~2: 
{\rm Constraints,~dimensional~reduction~conditions~and~section~conditions~of}~{\cal A}{\rm 5\mathchar`-brane.}&\nn
\eea
		These operators act on fields $\Phi(X)$ and $\Psi(X)$ as
\bea
&&\displaystyle\frac{\partial}{\partial X^M_0}\frac{\partial}{\partial X^N_0}\Phi(X_0)~=~0
		~=~\frac{\partial}{\partial X^M_0} \Phi(X_0)
		\frac{\partial}{\partial X^N_0}\Psi(X_0)\nn\\
&&\displaystyle\frac{\partial}{\partial \sigma_n}\frac{\partial}{\partial X^M}
		\Phi(\sigma, X(\sigma))~=~0~=~\frac{\partial}{\partial \sigma_n}\Phi(\sigma, X(\sigma))\frac{\partial}{\partial X^M}
		\Psi(\sigma, X(\sigma))~~~
\eea
where fields may be functions on $\sigma$ as $\Phi(\sigma, X(\sigma))$ and $\Psi(\sigma, X(\sigma))$.
		\par
		\vskip 6mm}	
	\item{${\cal M}$5-brane in 4-dimensional spacetime
		
		The dimensional reduction of the spacetime is obtained by solving the Virasoro constraint in \bref{SectionVira}  as
		\bea
		&P_{\underline{m}\underline{n}}(\sigma)=0~~,~~{\underline{m}=1,\cdots,4}~~\Rightarrow~
		\displaystyle\epsilon^{mm_1\cdots m_4}\frac{\partial}{\partial X^{m_1m_2}_0}P_{m_3m_4}(\sigma)=0	
		~~~.&
		\eea 
		This condition makes  $X^{\underline{m}_1\underline{m}_2}=y^{\underline{m}_1\underline{m}_2}$ to be non-dynamical {and reduced dimensionally}. The remaining spacetime is 4 dimensions $P_{5\underline{m}}=p_{\underline{m}}\neq 0$.

The 4-dimensional spacetime is described by the covariant derivative $\dd_{\underline{m}}$.
The 6-dimensional covariant derivative $\dd_{\underline{m}\underline{n}}$ is maintained to construct SL(5) current algebra
		\bea
		{\renewcommand{\arraystretch}{1.8}
			\left\{
			\begin{array}{ccl}
				\dd_{\underline{m}}&=&p_{\underline{m}}\\
				\dd_{\underline{m}_1\underline{m}_2}&=&-\epsilon_{\underline{m}_1\cdots \underline{m}_4}
				\partial^{\underline{m}_3}x^{\underline{m}_4}
			\end{array}\right.}
		\label{SL4ACA}
		\eea
		with $X^{5\underline{m}}=x^{\underline{m}}$ and $P_{5\underline{m}}=p_{\underline{m}}$ which is not confused with the 0-mode momentum.
		
		The SL(5) current algebra of ${\cal M}$5-brane is
		\bea
		{\renewcommand{\arraystretch}{1.8}
			\left\{
			\begin{array}{ccl}
				\left[\dd_{\underline{m}}(\sigma),\dd_{\underline{n}}(\sigma')\right]
				&=&0\\
				\left[\dd_{\underline{m}_1}(\sigma),\dd_{\underline{m}_2 \underline{m}_3}(\sigma')\right]
				&=&2i\epsilon_{\underline{m}_1\cdots \underline{m}_4}
				\partial^{\underline{m_4}}\delta^{(5)}(\sigma-\sigma')\label{SL5SL4}\\
				\left[\dd_{\underline{m}_1\underline{m}_2}(\sigma),\dd_{\underline{m}_3 \underline{m}_4}(\sigma')\right]
				&=&0
			\end{array}\right.~~~,}
		\eea
		where the last algebra forces to $\partial^5=0$.
		
		The Virasoro operators of ${\cal M}$5-brane are 
		\bea
		{\renewcommand{\arraystretch}{1.8}\left\{
			\begin{array}{ccl}
				{\cal S}^{\underline{m}}&=&
				\frac{1}{2}\partial^{[\underline{m}}x^{\underline{n}]}p_{\underline{n}}\\
				{\cal S}^5&=&\frac{1}{4}\epsilon_{\underline{m}_1\cdots \underline{m}_4}
			(\partial^{\underline{m}_1}x^{\underline{m}_2})
			(\partial^{\underline{m}_3}x^{\underline{m}_4})\\
				{\cal H}&=&\frac{1}{4}p_{\underline{m}}\hat{\eta}^{\underline{m}\underline{n}}p_{\underline{n}}
				+\frac{1}{16}(\partial^{[\underline{m}_1}x^{\underline{m}_2]})
				\hat{\eta}_{\underline{m}_1[\underline{n}_1}
				\hat{\eta}_{\underline{n}_2]\underline{m}_2}
		(\partial^{[\underline{n}_1}x^{\underline{n}_2]})\\
				{\cal U}_5&=&\partial^{\underline{n}}p_{\underline{n}}
			\end{array}
			\right.\label{VirasoroSL4}}~~~.
		\eea
		
These constraints lead to the following dimensional reductions and  section conditions.
\bea
&{\renewcommand{\arraystretch}{3.0}
\begin{array}{|l|c|c|}
	\hline
&{\rm Virasoro}:~ {\cal S}^{\underline{m}}&{\rm Gau\ss{}~law}:~{\cal U}_m\\
\hline
{\rm dimensional}~	{\rm reduction}
&\frac{1}{2}p_{\underline{m}}(\sigma)(\partial^{[\underline{m}}x^{\underline{n}]})
&
p_{\underline{m}}(\sigma)\partial^{\underline{m}}
\\\hline
{\rm section}~			{\rm condition}
&
{\rm none}
&\displaystyle
\partial^{\underline{m}}
\frac{\partial}{\partial x^{\underline{m}}_0}\\\hline
\end{array}
\label{GaussM5}}&\nn\\\nn\\
&{\rm Table}~3: 
{\rm Constraints,~dimensional~reduction~conditions~and~section~conditions~of}~{\cal M}{\rm 5\mathchar`-brane.}&\nn
\eea		
		\par\vskip 6mm
	}
	\item{${\cal T}$-string in 6-dimensional spacetime
		
		The dimensional reduction condition of the world-volume is obtained by solving the Gau\ss{} law constraint in \bref{sectionGau} as
		\bea
		&\partial^{\underline{n}}=0~,~P_{\underline{m}}(\sigma)=0~~~\Rightarrow~
		\partial^{n}P_{mn}(\sigma)=0~~~.&
		\eea 
		These conditions make $\partial^5=\frac{\partial}{\partial \sigma}\neq 0$ and $X^{\underline{m}}$ to be non-dynamical (constant). The remaining spacetime is 6 dimensional
		$P_{\underline{m}\underline{n}}
		=p_{\underline{m}\underline{n}}\neq 0$.

The 6-dimensional spacetime is described by the covariant derivative $\dd_{\underline{m}\underline{n}}$.
The 4-dimensional covariant derivative vanishes $\dd_{\underline{m}}=0$
		\bea
		\dd_{\underline{m}_1\underline{m}_2}&=&
		p_{\underline{m}_1\underline{m}_2}
		+\frac{1}{2}\epsilon_{\underline{m}_1\cdots\underline{m}_4}
		\partial^5 x^{\underline{m}_3\underline{m}_4}		
		\label{T1CA}
		\eea
		with $X^{\underline{m}\underline{n}}=x^{\underline{m}\underline{n}}$.
		
		The O(3,3) current algebra of ${\cal T}$-string is
		\bea
		\left[\dd_{\underline{m}_1\underline{m}_2}(\sigma),\dd_{\underline{m}_3 \underline{m}_4}(\sigma')\right]
		&=&2i\epsilon_{\underline{m}_1\cdots \underline{m}_4}
        \partial_\sigma
        \delta(\sigma-\sigma')
		\nn~~~\label{O6CA}
		\eea
		with $\partial^5=\partial_\sigma$.
		
		The Virasoro operators of ${\cal T}$-string are
		\bea
		{\renewcommand{\arraystretch}{1.8}\left\{
			\begin{array}{ccl}
				{\cal S}^5&=&\frac{1}{16}
				\dd_{\underline{m}_1\underline{m}_2}\epsilon^{\underline{m}_1\cdots \underline{m}_4}
				\dd_{\underline{m}_3\underline{m}_4}\\
				{\cal H}&=&\frac{1}{16}	\dd_{\underline{m}_1\underline{m}_2}
				\hat{\eta}^{\underline{m}_1[\underline{n}_1}
				\hat{\eta}^{\underline{n}_2]\underline{m}_2}	\dd_{\underline{n}_1\underline{n}_2}
			\end{array}
			\right.\label{VirasoroTst}}
		\eea
		with 
		${\cal S}^{\underline{m}}=0={\cal U}_m$.
		
The Virasoro constraint ${\cal S}^5=0$ leads to the following dimensional reduction and the section condition.
\bea
&{\renewcommand{\arraystretch}{3.0}
	\begin{array}{|l|c|c|}
		\hline
		&{\rm Virasoro}:~ {\cal S}^{\underline{m}}&{\rm Gau\ss{}~law}:~{\cal U}_m\\
		\hline
		{\rm dimensional}~	{\rm reduction}
		&p_{\underline{m}_1\underline{m}_2}(\sigma)\epsilon^{\underline{m}_1\cdots\underline{m}_4}
  \displaystyle\frac{\partial}
		{\partial x^{\underline{m}_3\underline{m}_4}_0 }
		&	{\rm none}
		\\\hline
		{\rm section}~			{\rm condition}
		&
		\displaystyle\frac{\partial}
	{\partial x^{\underline{m}_1\underline{m}_2}_0 }
	\epsilon^{\underline{m}_1\cdots\underline{m}_4}
	\frac{\partial}{\partial x^{\underline{m}_3\underline{m}_4}_0 }
		&	{\rm none}\\\hline
	\end{array}
	\label{VirasoroT2S}}&\nn\\\nn\\
&{\rm Table}~4: 
{\rm Constraints,~dimensional~reduction~conditions~and~section~conditions~of}~{\cal T}\mathchar`-{\rm string.}&\nn
\eea
		\par\vskip 6mm
	}	
	\item{$S$tring in 3-dimensional spacetime

\begin{enumerate}	
\item{		From ${\cal T}$-string to $S$tring

The dimensional reduction of the spacetime is obtained by solving the Virasoro constraint in \bref{VirasoroT2S}  as
		\bea
		&P_{\bar{m}\bar{n}}(\sigma)=0~~,~~{\bar{m}=1,2,3}~~\Rightarrow~
	P_{\underline{m}_1\underline{m}_2}(\sigma)\epsilon^{\underline{m}_1\cdots\underline{m}_4}\displaystyle\frac{\partial}
	{\partial x^{\underline{m}_3\underline{m}_4}_0 }
	=0~~~~.&
		\eea 
		This condition makes  $X^{\bar{m}_1\bar{m}_2}=y^{\bar{m}_1\bar{m}_2}$ to be non-dynamical (constant). The remaining spacetime is 3 dimensions $P_{4\bar{m}}=p_{\bar{m}}\neq 0$.}	
\item{	From ${\cal M}5$-brane to $S$tring
		
		The dimensional reduction condition of the world-volume is obtained by solving the Gau\ss{} law constraint in \bref{GaussM5} as
		\bea
		&\partial^{\bar{n}}=0~,~P_{4}(\sigma)=0~~~\Rightarrow~
		P_{\underline{m}}	\partial^{\underline{m}}=0
		~~~.&
		\eea 
		In the 4-dimensional spacetime $\partial^5$ is considered to be 0.
		These conditions make $\partial^4=\frac{\partial}{\partial \sigma}=\partial_\sigma\neq 0$ and $X^{54}$ to be non-dynamical (constant). The remaining spacetime is 3 dimensions 
		$P_{4\bar{m}}
		=p_{\bar{m}}\neq 0$.		
}	
	\end{enumerate}
		
The 3-dimensional spacetime is described by the covariant derivative $\dd_{4\bar{m}}$. 
The 3-dimensional covariant derivative vanishes $\dd_{\bar{m}_1\bar{m}_2}$
		\bea
		{\renewcommand{\arraystretch}{1.8}\left\{
			\begin{array}{ccl}
				\dd_{\bar{m}}&=&	p_{\bar{m}}\\
				\dd_{\bar{m}_1\bar{m}_2}&=&
				\epsilon_{\bar{m}_1\bar{m}_2\bar{m}}
			\partial_\sigma x^{\bar{m}}		
			\end{array}\right.}
		\label{F1CA}
		\eea
		with $X^{4\bar{m}}=x^{\bar{m}}$ and $\sigma^5=\sigma$ via ${\cal T}$-string and $X^{5\bar{m}}=x^{\bar{m}}$  and $\sigma^4=\sigma$ via ${\cal M}5$-brane.
		
		The GL(3) current algebra becomes 
		\bea
		\left[\dd_{\bar{m}_1}(\sigma),\dd_{\bar{m}_2 \bar{m}_3}(\sigma')\right]
		&=&i\epsilon_{\bar{m}_1\bar{m}_2 \bar{m}_3}	\partial_\sigma \delta(\sigma-\sigma')
		~~~.\label{GL3CA}
		\eea
This is equivalent to the O(D,D) current algebra which is given by  
$\dd_M=(\dd_{\bar{m}},~\dd^{\bar{m}})$ with $\dd^{\bar{m}}=\frac{1}{2}\epsilon^{\bar{m}\bar{m}_1\bar{m}_2}\dd_{\bar{m}_1\bar{m}_2}$ 
and  the O(D,D) invariant metric $\eta_{MN}=\epsilon_{\bar{m}_1\bar{m}_2\bar{m}_3}$ as
\bea
		\left[\dd_{M}(\sigma),\dd_{N}(\sigma')\right]
		&=&i\eta_{MN}	\partial_\sigma \delta(\sigma-\sigma')
		~~~.
		\eea

		The Virasoro operators become 
		\bea
		{\renewcommand{\arraystretch}{1.8}\left\{
			\begin{array}{ccl}
				{\cal S}&=& 
    p_{\bar{m}}\partial_\sigma x^{\bar{m}}~=~\frac{1}{2}\dd_{M}\eta^{MN}\dd_N\\
				{\cal H}&=&\frac{1}{2}p_{\bar{m}}\hat{\eta}^{\bar{m}\bar{n}} p_{\bar{n}}+\frac{1}{2}\partial_\sigma x^{\bar{m}}		\hat{\eta}_{\bar{m}\bar{n}}
				\partial_\sigma x^{\bar{n}}	~=~\frac{1}{2}\dd_M\hat{\eta}^{MN}\dd_N	
			\end{array}
			\right.\label{Virasorost}}
		\eea
		with the double Lorentz invariant metric $\hat{\eta}^{MN}$. 
	There are no further conditions of the Virasoro and the Gau\ss{} law constraints;  ${\cal S}^{\underline{m}}=0={\cal U}_m$.
		\par\vskip 6mm
	}
\end{enumerate}

\par
\vskip 6mm
\section{ ${\cal A}$5-brane Lagrangians}\label{section:3}

\subsection{${\cal A}$5-brane Lagrangian with SL(5) U-duality symmetry}

The SL(5) U-duality symmetry is manifestly realized by the ${\cal A}$5-brane.
The spacetime background is described by the vielbein which is a SL(5)/SO(5) coset element  $E_{m}{}^{a}$
satisfying
\bea
E_{{m}_1}{}^{{a}_1}E_{{m}_2}{}^{{a}_2}
E_{{m}_3}{}^{{a}_3}E_{{m}_4}{}^{{a}_4}
E_{{m}_5}{}^{{a}_5}
\epsilon_{{a}_1{a}_2{a}_3{a}_4{a}_5}
=\epsilon_{{m}_1{m}_2{m}_3{m}_4{m}_5}\label{SL5vielbein}
\eea
with ${m},{a}=1,\cdots,5$.
The background metrices with tensor indices are
\bea
G_{{m}{n}}&=&E_{{m}}{}^{{a}}
\hat{\eta}_{{a}{b}}
E_{{n}}{}^{{b}}\nn\\
G_{{m}_1{m}_2;{n}_1{n}_2}&=&
E_{m_1}{}^{{a}_1}E_{m_2}{}^{{a}_2}
\hat{\eta}_{{a}_1[{b}_1}
\hat{\eta}_{{b}_2]{a}_2}
E_{n_1}{}^{{b}_1}
E_{n_2}{}^{{b}_2}
~~~.
\eea

The selfdual and anti-selfdual currents in a flat background $	\don\circ{F}_{{\rm SD}/\overline{\rm SD}}{}^{mn}$
and in a curved background $F_{{\rm SD}/\overline{\rm SD}}{}^{ab}$ in terms of \bref{SL(5)A5current} are given as
\bea
&&{\renewcommand{\arraystretch}{1.8}
	\left\{\begin{array}{l}
		\don\circ{F}_{\rm SD}{}^{m_1m_2}=F_{\tau}^{m_1m_2}-\frac{1}{2}
		\epsilon^{m_1\cdots m_5}s_{m_3}F_{\sigma;m_4m_5}
		+g\hat{\eta}^{m_1n_1}\hat{\eta}^{m_2n_2}F_{\sigma;n_1n_2}\\	\don\circ{F}_{\overline{\rm SD}}{}^{m_1m_2}=F_{\tau}^{m_1m_2}-\frac{1}{2}
		\epsilon^{m_1\cdots m_5}s_{m_3}F_{\sigma;m_4m_5}
		-g \hat{\eta}^{m_1n_1}\hat{\eta}^{m_2n_2}F_{\sigma;n_1n_2}
	\end{array}\right. }\label{SL5SDflatcurrent}~~~\\
&&F_{{\rm SD}/\overline{\rm SD}}{}^{a_1a_2}=\don\circ{F}_{{\rm SD}/\overline{\rm SD}}{}^{m_1m_2}E_{m_1}{}^{a_1}E_{m_2}{}^{a_2}
\label{SL5SDcurvedcurrent}~~~
\eea
where $\hat{\eta}^{mn}$ becomes $G^{mn}$ in a curved background.
$g$ and $s_m$ are 5-brane world-volume vielbein fields which are introduced as Lagrange multipliers of Virasoro constraints.

The Lagrangian of the ${\cal A}$5-brane $L_{\rm SL(5)}$ is given \cite{Hatsuda:2023dwx} as
\bea
I_{\rm SL(5)}&=&\displaystyle\int d\tau d^5\sigma~L_{\rm SL(5)}\nn\\
L_{\rm SL(5)}&=&
\frac{1}{2}\phi F_{\rm SD}{}^{ab}F_{\overline{\rm SD}ab}
+\frac{1}{2}\bar{\phi} (F_{\overline{\rm SD}}{}^{ab})^2\nn\\&&
+\frac{1}{2}\lambda_{ab}F_{\overline{\rm SD}}{}^{ac}
F_{\overline{\rm SD}}{}^{b}{}_c
-\frac{1}{4}\epsilon_{a_1\cdots a_5} \lambda^{a_1}
F_{\overline{\rm SD}}{}^{a_2a_3}
F_{\overline{\rm SD}}{}^{a_4a_5}\nn\\
&=&
\frac{\phi}{4} \don\circ{F}_{\rm SD}{}^{m_1m_2}
G_{m_1m_2;m_3m_4}
\don\circ{F}_{\overline{\rm SD}}{}^{m_3m_4}
+\frac{\bar{\phi}}{8}
\don\circ{F}_{\overline{\rm SD}}{}^{m_1m_2}
G_{m_1m_2;m_3m_4}
\don\circ{F}_{\overline{\rm SD}}{}^{m_3m_4}
\nn\\
&&+\frac{1}{2}\don\circ{\lambda}_{mn}F_{\overline{\rm SD}}{}^{ml_1}
G_{l_1l_2}
\don\circ{F}_{\overline{\rm SD}}{}^{nl_2 }
+\frac{1}{8}\epsilon_{m_1\cdots m_5} \don\circ{\lambda}{}^{m_1}
\don\circ{F}_{\overline{\rm SD}}{}^{m_2m_3}
\don\circ{F}_{\overline{\rm SD}}{}^{m_4m_5}
\label{ALagwithSL5}~~~
\eea
with symmetric traceless tensors $\lambda_{\hat{m}\hat{n}}$'s.

\par\vskip 6mm

\subsection{World-volume covariant ${\cal A}$5-brane Lagrangian with SL(6) duality symmetry}

The $G$=SL(5) U-duality symmetry is enlarged to $A$=SL(6)
by cooperating with the 6-dimensional world-volume Lorentz covariance.
The SL(6)/SO(6) coset parameter includes not only the 
target space vielbein SL(5)/SO(5) but also 6 components of the world-volume vielbein.
The background vielbein $E_{\hat{m}}{}^{\hat{a}}\in$ SL(6)/SO(6) satisfies
\bea
E_{\hat{m}_1}{}^{\hat{a}_1}E_{\hat{m}_2}{}^{\hat{a}_2}
E_{\hat{m}_3}{}^{\hat{a}_3}E_{\hat{m}_4}{}^{\hat{a}_4}
E_{\hat{m}_5}{}^{\hat{a}_5}E_{\hat{m}_6}{}^{\hat{a}_6}
\epsilon_{\hat{a}_1\hat{a}_2\hat{a}_3\hat{a}_4\hat{a}_5\hat{a}_6}
=\epsilon_{\hat{m}_1\hat{m}_2\hat{m}_3\hat{m}_4\hat{m}_5\hat{m}_6}\label{SL6vielbein}
\eea
with $\hat{m},\hat{a}=0,1,\cdots,5$.

This SL(6) covariant vielbein \bref{SL6vielbein} includes 
the 5-brane world-volume vielbein fields $g$ and $s_m$ 
as
\bea
E_{\hat{m}}{}^{\hat{a}}&=&
{\renewcommand{\arraystretch}{1.8}
	\left(\begin{array}{cc}
		E_0{}^{\hat{0}}&E_{0}{}^{a}\\
		E_m{}^{\hat{0}}&E_{m}{}^a		
	\end{array}\right)=
	\left(\begin{array}{cc}
		\displaystyle\frac{1}{g}&0\\-\displaystyle\frac{s_m}{g}&E_m{}^a	
	\end{array}\right)}\label{SL6A5M22}~~~
\eea
with $\hat{m}=(0,m)$, $\hat{a}=(\hat{0},a)$ and $m,a=1\cdots,5$.
It is denoted that 
the $E_m{}^a$ component of  SL(6) vielbein \bref{SL6vielbein} is different from the SL(5) vielbein $E_m{}^a$ in \bref{SL6A5M22} up to the determinant factor.
The number of degrees of freedom of the SL(6) vielbein is sum of the spacetime vielbein and the world-volume vielbein as
\bea
(6^2-1)-\frac{6\times 5}{2}=\left( (5^2-1)-\frac{5\times 4}{2}\right)+6
~~~.
\eea
This is generalized for a $p$-brane of ${\cal A}$-theory symmetry with $A/L$ coset as
\bea
{\rm dim}~\displaystyle\frac{A}{L}={\rm dim}~\displaystyle\frac{G}{H}+(p+1)~~~.
\eea

The SL(6) covariant field strengths are given by a simple form; 
the one in a flat background $\don\circ{F}{}^{\hat{m}\hat{n}\hat{l}}$ ( the same as \bref{SL6F} )
and the one in a curved background ${F}{}^{\hat{a}\hat{b}\hat{c}}$
as 
\bea
\don\circ{F}{}^{\hat{m}\hat{n}\hat{l}}=\frac{1}{2}\partial^{[\hat{m}}X^{\hat{n}\hat{l}]}~~,~~
{F}{}^{\hat{a}\hat{b}\hat{c}}=\don\circ{F}{}^{\hat{m}\hat{n}\hat{l}}E_{\hat{m}}{}^{\hat{a}}E_{\hat{n}}{}^{\hat{b}}E_{\hat{l}}{}^{\hat{c}}~~~.
\label{SL6currents}
\eea
The selfdual and the anti-selfdual field strength \bref{SL5SDflatcurrent} and \bref{SL5SDcurvedcurrent} are written in terms of the SL(6) current \bref{SL6currents} 
 with $\partial^{\hat{0}} =\partial_\tau$ and $\epsilon^{\hat{0}12345}=1$ as
\bea
F_{{\rm SD}/\overline{\rm SD}}{}^{{a}_1{a}_2}&=&
g\left(F^{\hat{0}{a}_1{a}_2}\pm \frac{1}{6}
\epsilon^{\hat{0}{a}_1{a}_2}{}_{a_3a_4a_5}F^{a_3{a}_4{a}_5}\right)~~~.\label{SDaSDSL(6)}
\eea
Then the ${\cal A}$5-brane Lagrangian \bref{ALagwithSL5} is rewritten in terms of the SL(6) covariant field strength \bref{SL6currents}.
The world-volume covariant ${\cal A}$5-brane Lagrangian $L_{\rm SL(6)}$ is given \cite{Hatsuda:2023dwx}  as 
\bea
I_{\rm SL(6)}&=&\displaystyle\int d^6\sigma~L_{\rm SL(6)}\nn\\
L_{\rm SL(6)}&=&-\frac{1}{12}\Phi F^{\hat{a}_1\hat{a}_2\hat{a}_3}
F_{\hat{a}_1\hat{a}_2\hat{a}_3}
+\frac{1}{2}\Lambda_{\hat{a}\hat{b}}
F^{\hat{a}\hat{c}_1\hat{c}_2}
F^{\hat{b}}{}_{\hat{c}_1\hat{c}_2}
+\frac{1}{12}\epsilon_{\hat{a}_1\cdots \hat{a}_6}
\tilde{\Lambda}_{\hat{b}}{}^{\hat{a}_1}
F^{\hat{a}_2\hat{a}_3\hat{a}_4}
F^{\hat{a}_5\hat{a}_6\hat{b}}\nn\\
&=&-\frac{1}{72}\Phi \don\circ{F}{}^{\hat{m}_1\hat{m}_2\hat{m}_3}
G_{\hat{m}_1\hat{m}_2\hat{m}_3;\hat{m}_4\hat{m}_5\hat{m}_6}
\don\circ{F}{}^{\hat{m}_4\hat{m}_5\hat{m}_6}
+\frac{1}{8}\don\circ{\Lambda}{}_{\hat{m}\hat{n}}
\don\circ{F}{}^{\hat{m}\hat{l}_1\hat{l}_2}
G_{\hat{l}_1\hat{l}_2;\hat{l}_3\hat{l}_4}
\don\circ{F}{}^{\hat{n}\hat{l}_3\hat{l}_4}\nn\\
&&+\frac{1}{12}\epsilon_{\hat{m}_1\cdots \hat{m}_6}
\don\circ{\tilde{\Lambda}}{}_{\hat{n}}{}^{\hat{m}_1}
\don\circ{F}{}^{\hat{m}_2\hat{m}_3\hat{m}_4}
\don\circ{F}{}^{\hat{m}_5\hat{m}_6\hat{n}}\nn\\
\label{ALagwithSL6}
\eea
where $\Phi$, $\Lambda_{\hat{a}\hat{b}}$ are Lagrange multipliers  
with symmetric traceless tensors 
$\Lambda_{\hat{a}\hat{b}}$'s. 
The background metrices with tensor indices are
\bea
G_{\hat{m}_1\hat{m}_2;\hat{n}_1\hat{n}_2}&=&
E_{\hat{m}_1}{}^{\hat{a}_1}E_{\hat{m}_2}{}^{\hat{a}_2}
\hat{\eta}_{\hat{a}_1[\hat{b}_1}
\hat{\eta}_{\hat{b}_2]\hat{a}_2}
E_{\hat{n}_1}{}^{\hat{b}_1}
E_{\hat{n}_2}{}^{\hat{b}_2}\nn\\
G_{\hat{m}_1\hat{m}_2\hat{m}_3;\hat{n}_1\hat{n}_2\hat{n}_3}&=&
E_{\hat{m}_1}{}^{\hat{a}_1}E_{\hat{m}_2}{}^{\hat{a}_2}
E_{\hat{m}_3}{}^{\hat{a}_3}
\hat{\eta}_{\hat{a}_1[\hat{b}_1}
\hat{\eta}_{\hat{b}_2|\hat{a}_2|}
\hat{\eta}_{\hat{b}_3]\hat{a}_3}
E_{\hat{n}_1}{}^{\hat{b}_1}
E_{\hat{n}_2}{}^{\hat{b}_2}E_{\hat{n}_3}{}^{\hat{b}_3}
~~~.
\eea

\par
\vskip 6mm
\section{Lagrangian of D-dimensional \textbf{\textit{S}}tring from
	O(D,D) ${\cal T}$-string}\label{section:4}

In this section we derive the O(D,D) ${\cal T}$-string Lagrangian from the O(D,D) Hamiltonian by the double zweibein method \cite{Hatsuda:2018tcx,Hatsuda:2019xiz}.
Then the reduction procedure from the O(D,D) ${\cal T}$-string Lagrangian to the conventional string Lagrangian is presented. 
\par
\vskip 6mm
\subsection{O(D,D) ${\cal T}$-string}

We begin with the sigma model string Lagrangian
\bea
I&=&\displaystyle\int d^2\sigma~L\nn\\
L&=&-\frac{1}{2}\partial_\mu x^m(\sqrt{-h}h^{\mu\nu}g_{mn}+\epsilon^{\mu\nu}B_{mn})\partial_\nu x^n~~~\label{stringL}
\eea
with $\mu=(\tau,~\sigma)$.
In the conformal gauge the Lagrangian becomes
\bea
L
&=&\frac{1}{2}(\dot{x}^mg_{mn}\dot{x}^n-x'^mg_{mn}x'^n)-\dot{x}^mB_{mn}x'^n\nn\\
&=&\frac{1}{2}(\dot{x}^m~x'^m)
\left(\begin{array}{cc}g_{mn}&B_{mn}\\B_{mn}&g_{mn}	\end{array}	\right)
\left(\begin{array}{cc}1&0\\0&-1	\end{array}	\right)
\left(\begin{array}{c}\dot{x}^n\\x'^n	\end{array}	\right)\nn\\
&=&\frac{1}{2}\partial_+ x^m(g_{mn}+B_{mn})\partial_- x^n\label{string+-}~~~,
\eea 
with $\dot{x}=\partial_\tau x$,  ${x}'=\partial_\sigma x$ and  $\partial_\pm x=\dot{x}\pm x'$.

The Hamiltonian is given by the Legendre transformation where the canonical momentum of $x^m$ is given by $p_m={\partial L}/{\partial \dot{x}^m}$,
\bea
H&=&p_m\dot{x}^m-L\nn\\
&=&\frac{1}{2}(p_m~x'^m)
\left(\begin{array}{cc}g^{mn}&g^{ml}B_{ln}\\-B_{ml}g^{ln}&g_{mn}-B_{ml}g^{lk}B_{kn}	\end{array}	\right)
\left(\begin{array}{c}p_n\\x'^n	\end{array}	\right)\label{Hamstring}\\
&=&\frac{1}{2
}\{(p_m-x'^lB_{lm})g^{mn}(p_n+B_{nk}x'^k)+x'^mg_{mn}x'^n
\}\nn~~.
\eea
The background field is the O(D,D) matrix $G^{MN}$ written in terms of the vielbein $E_{A}{}^M$ as $E_A{}^M\to h_A{}^B~E_B{}^N{}g_N{}^M $, $h\in$SO(D$-$1,1) and $g\in$ O(D,D)
\bea
E_{A}{}^M\eta_{MN} E_{B}{}^N=\eta_{AB}~~~.\label{Orthogonal}
\eea
The background metric $G^{MN}$ in the string Hamiltonian \bref{Hamstring} is given as  
\bea
G^{MN}&=&
\left(\begin{array}{cc}g^{mn}&g^{ml}B_{ln}\\-B_{ml}g^{ln}&g_{mn}-B_{ml}g^{lk}B_{kn}	\end{array}	\right)~=~E_A{}^M\hat{\eta}^{AB}E_B{}^N~~\\
E_A{}^M&=&\left(\begin{array}{cc}e_a{}^m&e_a{}^lB_{lm}\\0&e_m{}^a	\end{array}	\right)~~~,\nn
\eea
while its inverse is given by
\bea
G_{MN}&=&\left(\begin{array}{cc}g_{mn}-B_{ml}g^{lk}B_{kn}&-B_{ml}g^{ln}\\g^{ml}B_{ln}&g^{mn}	\end{array}	\right)~=~
E_M{}^A\hat{\eta}_{AB}E_N{}^B~~~\label{ODDmetric}\\
E_M{}^A&=&\left(\begin{array}{cc}e_m{}^a&-B_{ml}e_a{}^l\\0&e_a{}^m	\end{array}	\right)~~~.\nn
\eea
This  O(D,D) background metric is utilized in the Lagrangian with manifest O(D,D) T-duality symmetry.

The O(D,D) covariant space is constructed in such a way that the O(D,D) covariant derivative $\dd_M(\sigma)$ algebra  satisfies the same algebra of $\dd_M=(p_m,x'^m)$
up to the normalization
\bea
[\dd_M(\sigma),\dd_N(\sigma')]&=&2i\eta_{MN}\partial_\sigma \delta(\sigma-\sigma')~~,~~\eta_{MN}=
\left(\begin{array}{cc}&\delta_m^n\\\delta_n^m&	\end{array}	\right)~~~.\label{CAstring}
\eea
The covariant derivative $\dd_M$ is realized in terms of the doubled coordinate $X^M$ and $P_M$ with $[P_M(\sigma),X^N(\sigma')]=-i\delta_M^N\delta(\sigma-\sigma')$ as
\bea
\don\circ{\dd}_M~=~P_M+\partial_\sigma X^N\eta_{NM}~~~,
\eea 
which is left moving current in the doubled space. 
The right moving current is also introduced as
\bea
\tilde{\dd}_M~=~P_M-\partial_\sigma X^N\eta_{NM}~~~
\eea 
which satisfies the same current algebra \bref{CAstring} with opposite sign.
The number of canonical variables of the doubled space are 4D, while the physical one is 2D.
The 2D equations $\tilde{\dd}_M=0$ is the  usual selfduality condition to suppress 2D unphysical degrees of freedom, so we call $\tilde{\dd}_M$ ``anti-selfdual current".
Another current $\don\circ{\dd}_M$ is selfdual current.

There are two sets of Virasoro operatros written in terms of the selfdual current and  the anti-selfdual current
\bea
\left\{\begin{array}{ccl}
	{\cal H}&=&\frac{1}{4}\don\circ{\dd}_{M}\hat{\eta}^{MN}\don\circ{\dd}_N\\
	{\cal S}&=&\frac{1}{4}\don\circ{\dd}_{M}\eta^{MN}\don\circ{\dd}_N
\end{array}
\right.~~,~~
\left\{\begin{array}{ccl}
	\tilde{\cal H}&=&\frac{1}{4}\tilde{\dd}_{M}\hat{\eta}^{MN}\tilde{\dd}_N\\
	\tilde{\cal S}&=&\frac{1}{4}\tilde{\dd}_{M}\eta^{MN}\tilde{\dd}_N
\end{array}
\right.~~.
\eea
${\cal H}$ and ${\cal S}$ satisfy the Virasoro algebra 
\bea
[{\cal S}(\sigma),{\cal S}(\sigma')\}]&=&i\{{\cal S}(\sigma)+{\cal S}(\sigma')\}\partial_\sigma\delta(\sigma-\sigma')\nn\\
\left[{\cal S}(\sigma),{\cal H}(\sigma')\}\right]&=&i\{{\cal H}(\sigma)+{\cal H}(\sigma')\}\partial_\sigma\delta(\sigma-\sigma')\\
\left[{\cal H}(\sigma),{\cal H}(\sigma')\}\right]&=&i\{{\cal S}(\sigma)+{\cal S}(\sigma')\}\partial_\sigma\delta(\sigma-\sigma')\nn~~~,
\eea
while
$\tilde{\cal H}$ and $\tilde{\cal S}$ satisfy the same Virasoro algebra with opposite signs on the right hand side.

As seen in the Hamiltonian in curved background \bref{Hamstring}  currents $\dd_M$  coupled to the vielbein as
\bea
\dd_A=E_A{}^M\don\circ{\dd}_M~~,~~\tilde{\dd}_A=E_A{}^M\tilde{\dd}_M~~~.\eea
In curved background the Virasoro constraints become
\bea
\left\{\begin{array}{ccl}
	{\cal H}&=&\frac{1}{4}{\dd}_{A}\hat{\eta}^{AB}{\dd}_B=\frac{1}{4}\don\circ{\dd}_{M}G^{MN}\don\circ{\dd}_N
	\\
	{\cal S}&=&\frac{1}{4}{\dd}_A\eta^{AB}{\dd}_B=\frac{1}{4}\don\circ{\dd}_{M}\eta^{MN}\don\circ{\dd}_N
\end{array}
\right.~~,~~
\left\{\begin{array}{ccl}
	\tilde{\cal H}&=&\frac{1}{4}\tilde{\dd}_{A}\hat{\eta}^{AB}\tilde{\dd}_B=\frac{1}{4}\tilde{\dd}_{M}G^{MN}\tilde{\dd}_N \\
	\tilde{\cal S}&=&\frac{1}{4}\tilde{\dd}_{A}\eta^{AB}\tilde{\dd}_B=\frac{1}{4}\tilde{\dd}_{M}\eta^{MN}\tilde{\dd}_N
\end{array}
\right.~.
\eea

The O(D,D) covariant Hamiltonian is given by the sum of all these Virasoro constraints with Lagrange multipliers which are doubled zweibeins \cite{Hatsuda:2018tcx}
\bea
H&=&g{\cal H}+s{\cal S}+\tilde{g}\tilde{\cal H}+\tilde{s}\tilde{\cal S}\nn\\
&=&\frac{1}{2}\left[
P_A M^{AB} P_B+2P_A N^{AC}\eta_{CB}X'^B+X'^A\eta_{AC}M^{CD}\eta_{DB}X'^B
\right]\label{HamODD}
\eea
with $P_A=P_ME_A{}^M$ and $X'^A\equiv X'^ME_M{}^A$.
We used the fact that the covariant derivatives are rewritten as
$ \dd_A=P_A+X'^B\eta_{BA}$ and $ \tilde{\dd}_A=P_A-X'^B\eta_{BA}$
by the orthogonal condition \bref{Orthogonal}.
Matrices $M^{AB}$ and $N^{AB}$ are given as
\bea
M^{AB}&=&\frac{g+\tilde{g}}{2}\hat{\eta}^{AB}+\frac{s+\tilde{s}}{2}\eta^{AB}\nn\\
N^{AB}&=&\frac{g-\tilde{g}}{2}\hat{\eta}^{AB}+\frac{s-\tilde{s}}{2}\eta^{AB}~~~,\label{MN}
\eea
with the inverse of $M^{AB}$ as
\bea
M^{-1}{}_{AB}&=&\frac{2}{(g+\tilde{g})^2-(s+\tilde{s})^2}
\left\{(g+\tilde{g})\hat{\eta}_{AB}-(s+\tilde{s})\eta_{AB}
\right\}~~~.
\eea

The Legendre transformation of the Hamiltonian \bref{HamODD} with \bref{MN} leads to the following Lagrangian
\bea
L&=&P_M\dot{X}^M-H~=~\frac{1}{2}J_+{}^AM^{-1}{}_{AB}J_-{}^B\nn\\
&&\left\{
\begin{array}{ccl}
	J_+{}^A&=&\dot{X}^A+(\tilde{g}\hat{\eta}^{AB}+\tilde{s}\eta^{AB})
	\eta_{BC}X'^C\\
	J_-{}^A&=&\dot{X}^A-({g}\hat{\eta}^{AB}+{s}\eta^{AB})
	\eta_{BC}X'^C
\end{array}
\right.\label{Jst+-}\label{Lagst+-}
\eea
with $\dot{X}^A\equiv \dot{X}^ME_M{}^A$.

The  Lagrangian in \bref{Lagst+-} can be written in terms of the selfdual current and the anti-selfdual  current  which is equal to $J_-$ in \bref{Jst+-}. 
The selfdual and anti-selfdual currents are given by
\bea
\left\{
\begin{array}{ccl}
	J_{\rm SD}{}^A&=&(\dot{X}^A-sX'^A)+{g}\hat{\eta}^{AB}
	\eta_{BC}X'^C\\
	J_{\overline{\rm SD}}{}^A&=&(\dot{X}^A-sX'^A)-{g}\hat{\eta}^{AB}
	\eta_{BC}X'^C
\end{array}
\right.\label{JstSDASD}~~~.
\eea
The selfdual and anti-selfdual currents in the flat background,  $J_{\rm SD/\overline{SD}}{}^M=J_{\rm SD/\overline{SD}}{}^AE_A{}^M$, are written as
\bea
\left\{
\begin{array}{ccl}
	\don\circ{J}_{\rm SD}{}^M&=&(\dot{X}^M-sX'^M)+{g}\hat{\eta}^{MN}
	\eta_{NL}X'^L\\
	\don\circ{J}_{\overline{\rm SD}}{}^M&=&(\dot{X}^M-sX'^M)-{g}\hat{\eta}^{MN}\eta_{NL}X'^L
\end{array}
\right.\label{JstSDASDflat}~~~.
\eea
It is denoted that $\hat{\eta}^{MN}$ becomes $G^{MN}$ in a curved background.
The resultant O(D,D) covariant Lagrangian for a 
${\cal T}$-string is given \cite{Hatsuda:2018tcx} as
\bea
I&=&\displaystyle\int d\tau d\sigma~L\nn\\
L&=&\phi J_{\rm SD}{}^A\hat{\eta}_{AB}J_{\overline{\rm SD}}{}^B
+\bar{\phi} J_{\overline{\rm SD}}{}^A\hat{\eta}_{AB}J_{\overline{\rm SD}}{}^B
+\tilde{\phi} J_{\overline{\rm SD}}{}^A{\eta}_{AB}J_{\overline{\rm SD}}{}^B\nn\\
&=&\phi \don\circ{J}_{\rm SD}{}^MG_{MN}\don\circ{J}_{\overline{\rm SD}}{}^N
+\bar{\phi} \don\circ{J}_{\overline{\rm SD}}{}^MG_{MN}\don\circ{J}_{\overline{\rm SD}}{}^N
+\tilde{\phi} \don\circ{J}_{\overline{\rm SD}}{}^M{\eta}_{MN}\don\circ{J}_{\overline{\rm SD}}{}^N\label{LagODDst}
~~~.
\eea
The first term is the kinetic term, while the rest are constraints that are squares of the anti-selfdual currents.
The Lagrange multipliers $\phi$, $\bar{\phi}$ and $\tilde{\phi}$ are
related to the doubled zweibeins as
\bea
\left\{\begin{array}{ccl}
	\phi&=&\displaystyle\frac{1}{2g}\\
	\bar{\phi}&=&\displaystyle\frac{1}{2g[(g+\tilde{g})^2-(s+\tilde{s})^2]}
	\left\{(s+\tilde{s})^2+g^2-\tilde{g}^2\right\}\\
	\tilde{\phi}&=&-\displaystyle\frac{s+\tilde{s}}{(g+\tilde{g})^2-(s+\tilde{s})^2}
\end{array}
\right.
\eea

\par
\vskip 6mm
\subsection{ ${S}$tring from O(D,D) ${\cal T}$-string}\label{Reduction2st}

We break the O(D,D) T-duality symmetry of ${\cal T}$-string into the GL(D) symmetry of the usual string.
The background gauge field of ${\cal T}$-string is O(D,D)/O(D$-$1,1)$^2$ coset parameter which includes the D-dimensional metric $g_{mn}$ and $B_{mn}$ field,
while the background gauge field of a string is  GL(D)/SO(D-1,1) coset parameter which includes only $g_{mn}$.
In this subsection we use the coordinate  
$X^M=(x^m,~y_m)$ with off-diagonal $\eta_{MN}$ to describe ${\cal T}$-string,
while the left/right moving coordinate with diagonal 
$\eta_{MN}=({\bf 1},-{\bf 1})$ was used in the reference \cite{Hatsuda:2018tcx}.   
The Weyl/Lorentz gauge of the zweibein \cite{Hatsuda:2018tcx} is given as
\bea
\varepsilon_\pm{}^\mu
=\left(\begin{array}{cc}
	\varepsilon_+{}^\tau&\varepsilon_+{}^\sigma\\
	\varepsilon_-{}^\tau&\varepsilon_-{}^\sigma
\end{array}\right)=
\left(\begin{array}{cc}
	1&g-s\\
	1&-g-s
\end{array}\right)~
~~~.\label{zweibein}
\eea
The left/right moving modes with the zweibein  is   $\varepsilon_\pm X\equiv \varepsilon_\pm{}^\mu \partial_\mu X$.
The selfdual and anti-selfdual currents \bref{JstSDASDflat} are expressed as
\bea
\left\{\begin{array}{lcl}
	\don\circ{J}_{{\rm SD}/\overline{\rm SD}}{}^{{m}}&=&\dot{x}^{{m}}-sx'^{{m}}
	\pm g \hat{\eta}^{{m}{n}}y'_{{n}}~=~\varepsilon_\tau x^{{m}}\pm g\hat{\eta}^{{m}{n}}\varepsilon_\sigma y_{{n}}\\
	\don\circ{J}_{{\rm SD}/\overline{\rm SD}}{}_{;{m}}
	&=&\dot{y}_{{m}}-sy'_{{m}}
	\pm g 	\hat{\eta}_{{m}{n}} 	x'^{{n}}~
	=~\varepsilon_\tau  y_{{m}}  \pm  g\hat{\eta}_{{m}{n}}
	\varepsilon_\sigma x^{{n}}
\end{array}\right.~~~.\label{sdasdxy}
\eea 
with
\bea
\varepsilon_\mu X\equiv \varepsilon_\mu{}^\nu \partial_\nu X~~,~~
\varepsilon_\mu{}^\nu
=\left(\begin{array}{cc}
	\varepsilon_\tau{}^\tau&\varepsilon_\tau{}^\sigma\\
	\varepsilon_\sigma{}^\tau&\varepsilon_\sigma{}^\sigma
\end{array}\right)=
\left(\begin{array}{cc}
	1&-s\\
	0&1
\end{array}\right)
\eea

The condition of vanishing the anti-selfdual current  in $s=0$ and $g=1$ gauge leads to the selfduality constraint in flat space as $\partial_\mu{y}=\epsilon_{\mu\nu}\partial^\nu x$ with $\partial^{\tau}=-\partial_\tau$.
In the gauge $\phi=\frac{1}{2g}$ and $\bar{\phi}=0=\tilde{\phi}$,
corresponding to $g=\tilde{g}$ and $s+\tilde{s}=0$, the O(D,D) covariant Lagrangian \bref{LagODDst}
is written as
\bea
&&\displaystyle\frac{1}{2g}~ J_{\rm SD}{}^A\hat{\eta}_{AB}J_{\overline{\rm SD}}{}^B\nn
\\&&
~~~=~\displaystyle\frac{1}{2g}(\dot{X}-sX')^A\hat{\eta}_{AB}(\dot{X}-sX')^B
-\displaystyle\frac{g}{2}X'^C\eta_{CD}\hat{\eta}^{DA}\hat{\eta}_{AB}\hat{\eta}^{BF}\eta_{FG}X'^G\nn\\
&&~~~=~\displaystyle\frac{1}{2g}(\dot{X}-sX')^ME_M{}^A\hat{\eta}_{AB}E_N{}^B(\dot{X}-sX')^N
-\displaystyle\frac{g}{2}X'^M\eta_{ML}E_{A}{}^L\hat{\eta}^{AB}E_B{}^K\eta_{KN}X'^N\nn\\
&&~~~=~\displaystyle\frac{1}{2g}~\varepsilon_+X^{M}~ G_{MN} ~\varepsilon_- X^N\nn\\
&&~~~=~\displaystyle\frac{1}{2g}~\varepsilon_+X^{M}E_M{}^A~ \hat{\eta}_{AB} ~E_N{}^B\varepsilon_- X^N
~~~.\label{kineticTst}
\eea
The orthogonality condition is used in the second equality,
 $\eta_{ML}G^{LK}\eta_{KN}=G_{MN}$, so 
$E_M{}^A\eta_{AB}=\eta_{MN}E_B{}^N$,
is used in the last equality.
 In terms of $x^m,y_m$ coordinates it is given by 
\bea
&&\displaystyle\frac{1}{2g}~ J_{\rm SD}{}^A\hat{\eta}_{AB}J_{\overline{\rm SD}}{}^B\nn\\
&&~~~=~\displaystyle\frac{1}{2g}~(\varepsilon_+ x^m~\varepsilon_+y_m)
\left(\begin{array}{cc}g_{mn}-B_{ml}g^{lk}B_{kn}&-B_{ml}g^{ln}\\g^{ml}B_{ln}&g^{mn}	\end{array}	\right)
\left(
\begin{array}{c}
	\varepsilon_- x^n \\\varepsilon_- y_n
\end{array}
\right)\label{kinetic}\label{ODDxy}\\
&&~~~=~\displaystyle\frac{1}{2g}~(\varepsilon_+ x^m~\varepsilon_+y_m)
\left(\begin{array}{cc}e_m{}^a&-B_{ml}e_a{}^l\\0&e_a{}^m\end{array}	\right)
\left(\begin{array}{cc}\eta_{ab}&0\\0&\eta^{ab}\end{array}	\right)
\left(\begin{array}{cc}e_n{}^b&0\\-B_{nk}e_b{}^k&e_b{}^n\end{array}	\right)
\left(
\begin{array}{c}
	\varepsilon_- x^n \\\varepsilon_- y_n
\end{array}
\right)\nn\\
&&~~~=~\displaystyle\frac{1}{2g}~\left[\varepsilon_+x^m g_{mn}\varepsilon_- x^n
+(\varepsilon_+y_m - \varepsilon_+x^lB_{lm})g^{mn}
(\varepsilon_-y_n + B_{nk}\varepsilon_-x^k)\right]\nn
\eea
We break the O(D,D) symmetry into the GL(D) symmetry 
by the dimensional reduction \bref{dimreduction}.
The resultant Lagrangian is the kinetic term of the usual string with the zweibein field;
\bea
L_0&=&\displaystyle\frac{1}{2g}~ \varepsilon_+x^m~ g_{mn}~ \varepsilon_- x^n~~~\label{kinusualst}
\eea

In order to obtain the Wess-Zumino term we add the total derivative term 
\bea
\partial_\mu(\epsilon^{\mu\nu}x^m\partial_\nu y_m)=\dot{x}y'-x'\dot{y}=-\frac{1}{2g}(\varepsilon_+ x ~\varepsilon_-y -\varepsilon_+y ~\varepsilon_- x) \label{totalderst}
\eea
to the O(D,D) Lagrangian ${L}$ \bref{ODDxy} 
\bea
&&\frac{1}{2g} J_{\rm SD}{}^A\hat{\eta}_{AB}J_{\overline{\rm SD}}{}^B
-\partial_\mu(\epsilon^{\mu\nu}x^m\partial_\nu y_m)\label{stringLagWZ}\\
&&~~~=
\frac{1}{2g}
\left\{
\varepsilon_+x^m g_{mn} \varepsilon_ -^nx \right.\nn\\
&&~~~~~\left.+(\varepsilon_+ y_m -\varepsilon_+ x^l B_{lm}-\varepsilon_+ x^l g_{lm}) g^{mn} (\varepsilon_- y_n+B_{nk}\varepsilon_- x^k+g_{nk}\varepsilon_- x^k)\right.\nn\\
&&~~~~~\left.+\varepsilon_+x^m g_{mn} \varepsilon_- x^n+2\varepsilon_+x^m B_{mn} \varepsilon_- x^n\right\}~~~\nn
\eea
By the dimensional reduction \bref{dimreduction}
the Lagrangian with the total derivative term reduces into the string Lagrangian in curved background with the Wess-Zumino term 
as the curved world-sheet version of \bref{string+-},
\bea
L_0+L_{\rm WZ}&=&\frac{1}{g}~\varepsilon_+x^m (g_{mn}+B_{mn}) \varepsilon_- x^n~~~.\label{GLDstLag}
\eea
The zweibeins in \bref{GLDstLag} and \bref{stringL} are related as
\bea
g=-\displaystyle\frac{2}{\sqrt{-h}h^{00}}~~,~~
s=-\displaystyle\frac{h^{01}}{h^{00}}
~~~.
\eea 

\par
\vskip 6mm
\section{Lagrangians of \textbf{\textit{S}}tring  via 
	${\cal T}$-string  from ${\cal A}$5-brane}\label{section:5}

In this section, we derive the ${\cal T}$-string Lagrangian from the ${\cal A}5$-brane Lagrangian.
The resulting ${\cal T}$-string Lagrangian is formulated in terms of an SL(4) rank-two antisymmetric tensor coordinate, which is coupled to the string background.
We then present the reduction procedure from the ${\cal T}$-string Lagrangian to the conventional string Lagrangian.
\par
\vskip 6mm

\subsection{${\cal T}$-string from ${\cal A}$5-brane}\label{section:5-1}

The O(3,3) ${\cal T}$-string from ${\cal A}$5-brane is 
described by  the SL(4) rank-two anti-symmetric tensor 
coordinate $X^{\underline{mn}}=(x^{\bar{m}},~y^{\bar{m}\bar{n}})$ with $\underline{m}=1,\cdots, 4$ and ~$\bar{m}=1,2,3$ as listed in \bref{rep}.
The SL(6) rank-two tensor coordinate is decomposed as SL(6) $\to$ SL(5) $\to$ SL(4)   as 
$X^{\hat{m}\hat{n}}=(X^{0n}=Y^n,~X^{mn})$ $\to$
$X^{{m}{n}}=(X^{5\underline{n}}=Y^{\underline{m}},~X^{\underline{m}\underline{n}})$ $\to$ $X^{\underline{m}\underline{n}}=(X^{4\bar{m}}=x^{\bar{m}},~X^{\bar{m}\bar{n}}=y^{\bar{m}\bar{n}})$ with $\hat{m}=0,1,\cdots,5$ and $m=1,\cdots,5$.
The 6-dimensional world-volume derivative is reduced into the string world-sheet derivatives as $\partial^{\hat{m}}=(\partial^0=\partial_\tau,~\partial^5=\partial_\sigma,~\partial^{\underline{m}}=0)$. 
The SL(6) field strength for the ${\cal T}$-string has the following components
\bea
&&\don\circ{F}{}^{0\underline{mn}}~=~\partial_\tau X^{\underline{mn}}~,~~
\don\circ{F}{}^{5\underline{mn}}~=~\partial_\sigma X^{\underline{mn}}\label{SL6FSTst}~~,~~\don\circ{F}{}^{05\underline{m}}~=~ 0~=~
 \don\circ{F}{}^{\underline{mnl}} ~~~.
\eea

The SL(6) vielbein for the ${\cal T}$-string 
has a block diagonal form as
\bea
E_{\hat{m}}{}^{\hat{a}}&=&
{\renewcommand{\arraystretch}{1.8}
	\left(\begin{array}{ccc}
		E_0{}^{\hat{0}}&E_{0}{}^{\hat{5}}&E_{0}{}^{\underline{a}}\\
		E_5{}^{\hat{0}}&E_{5}{}^{\hat{5}}&E_{5}{}^{\underline{a}}\\
		E_{\underline{m}}{}^{\hat{0}}&E_{\underline{m}}{}^{\hat{5}}&E_{\underline{m}}{}^{\underline{a}}		
	\end{array}\right)=
	\left(\begin{array}{ccc}
		\displaystyle\frac{1}{g}&0&0\\
				-\displaystyle\frac{s}{g}&1&0\\
		~~0~~&~~0~~&g^{1/4}E_{\underline{m}}{}^{\underline{a}}	
	\end{array}\right)}
\label{TstinA5}~~~.
\eea
The selfdual and the anti-selfdual currents are the following combinations of the SL(6) field strengths in \bref{SL6FSTst} with \bref{SL6currents} as
\bea
{J}_{{\rm SD}/\overline{\rm SD}}{}
^{\underline{a}_1\underline{a}_2}&=&g\left( F^{\hat{0}\underline{a}_1\underline{a}_2}
\pm \frac{1}{2} \epsilon^{\hat{0}\underline{a}_1\underline{a}_2}{}_{5\underline{a}_3\underline{a}_4} 
F^{\hat{5}\underline{a}_3\underline{a}_4}\right)~~~.
\eea
The  zweibein fields $g$ and $s$ 
are part of the SL(6) vielbein  \bref{TstinA5} in the new SL(6) duality symmetry formulation in \bref{SL6currents},
contrast to that the world-volume vielbein fields are separated from the  SL(4) spacetime vielbein 
$E_{\underline{m}}{}^{\underline{a}}$ in the SL(5) formulation in \bref{SL5SDflatcurrent} and \bref{SL5SDcurvedcurrent}  as
\bea
{J}_{{\rm SD}/\overline{\rm SD}}{}
^{\underline{a}_1\underline{a}_2}&=&\don\circ{J}_{{\rm SD}/\overline{\rm SD}}{}
^{\underline{m}_1\underline{m}_2}E_{\underline{m}_1}{}^{\underline{a}_1}
E_{\underline{m}_2}{}^{\underline{a}_2}
\nn\\
\don\circ{J}_{{\rm SD}/\overline{\rm SD}}{}
^{\underline{m}_1\underline{m}_2}&=&
\varepsilon_\tau{X}^{\underline{m}_1\underline{m}_2}
\pm  \frac{1}{2}\hat{\eta}^{\underline{m}_1\underline{n}_1}
\hat{\eta}^{\underline{m}_2\underline{n}_2}
(-\epsilon_{\underline{n}_1\cdots \underline{n}_4})\varepsilon_\sigma X^{\underline{n}_3\underline{n}_4} 
%
 \label{SDaSDCSL4}~~~
\eea 
with \bref{zweibein}.
The minus sign in the last equation is  caused from $\epsilon_{\bar{m}_1\bar{m}_2\bar{m}_3}x^{\bar{m}_3}=-\epsilon_{\bar{m}_1\bar{m}_24\bar{m}_3}X^{4\bar{m}_3}$.
The O(3,3) invariant metric $\eta_{MN}$ becomes SL(4) invariant metric $\epsilon_{\underline{m}_1\cdots \underline{m}_4}$.
The current in \bref{SDaSDCSL4} is written in terms of $x^{\bar{m}}$ and $y^{\bar{m}\bar{n}}$ as
\bea
\left\{\begin{array}{lcl}
	\don\circ{J}_{{\rm SD}/\overline{\rm SD}}{}^{4\bar{m}}&=&\dot{x}^{\bar{m}}-sx'^{\bar{m}}
	\pm g  \frac{1}{2}\epsilon^{\bar{m}}{}_{\bar{n}_1\bar{n}_2}y'^{\bar{n}_1\bar{n}_2}
~=~\varepsilon_\tau x^{\hat{m}}\pm \frac{1}{2}\epsilon^{\hat{m}}{}_{\hat{n}_1\hat{n}_2}\varepsilon_\sigma y^{\hat{n}_1\hat{n}_2}
\\
	\don\circ{J}_{{\rm SD}/\overline{\rm SD}}{}^{\bar{m}_1\bar{m}_2}&=&\dot{y}^{\bar{m}_1\bar{m}_2}-sy'^{\bar{m}_1\bar{m}_2}
	\pm g\epsilon^{\bar{m}_1\bar{m}_2}{}_{\bar{n}}x'^{\bar{n}}~=~\varepsilon_\tau y^{\hat{m}_1\hat{m}_2}\pm \epsilon^{\hat{m}_1\hat{m}_2}{}_{\hat{n}}\varepsilon_\sigma x^{\hat{n}}
\end{array}\right.~~~
\eea 
which is related to the O(D,D) vector currents \bref{sdasdxy} with $y^{\bar{m}\bar{n}}\equiv\epsilon^{\bar{m}\bar{n}\bar{l}}y_{\bar{l}}$.

In order to obtain the usual 3-dimensional string Lagrangian we express  the spacetime vielbein $E_{\underline{m}}{}^{\underline{a}}$ 
$\in$ SL(4)/SO(4) in terms of the 3-dimensional 
metric $g_{\bar{m}\bar{n}}$ and the $B_{\bar{m}\bar{n}}$ field.
The O(3,3) vector index contraction and the SL(4) tensor index contraction are assumed to be equal up to the normalization as
\bea
dX^ME_M{}^A&=&dx^{\bar{m}}E_{\bar{m}}{}^A+dy_{\bar{m}}E^{\bar{m};A}
=dX^{4\bar{x}}E_{4\bar{m}}{}^A+\frac{1}{2}dX^{\bar{m}\bar{n}}E_{\bar{m}\bar{n}}{}^{A}=\frac{1}{2}
dX^{\underline{m}\underline{n}}E_{\underline{m}\underline{n}}{}^{A}~~~.\nn\\
\eea
We rewrite the O(D,D) vielbein in \bref{ODDmetric} in terms of tensor indices for D=3 case as
\bea
E_M{}^A&=&
\left(\begin{array}{cc}e_{\bar{m}}{}^{\bar{a}}&-B_{\bar{m}\bar{l}}e_{\bar{b}}{}^{\bar{l}}\epsilon^{\bar{a}_1\bar{a}_2\bar{b}}\\0&\epsilon_{\bar{m}_1\bar{m}_2\bar{m}}e_{\bar{b}}{}^{\bar{m}}\epsilon^{\bar{a}_1\bar{a}_2\bar{b}}	\end{array}	\right)~
\label{v2t}\\
&=&c
\left(\begin{array}{cc}E_{4\bar{m}}{}^{4\bar{a}}&E_{4\bar{m}}{}^{\bar{a}_1\bar{a}_2}\\E_{\bar{m}_1\bar{m}_2}{}^{4\bar{a}}
	&E_{\bar{m}_1\bar{m}_2}{}^{\bar{a}_1\bar{a}_2}	\end{array}	\right)
~=~cE_{\underline{m}_1\underline{m}_2}{}^{\underline{a}_1\underline{a}_2}~=~
cE_{[\underline{m}_1}{}^{\underline{a}_1}
E_{\underline{m}_2]}{}^{\underline{a}_2}
~~\nn
\eea
with a normalization factor $c$.
The vielbein with the tensor indices can be written as the product of the one with the vector indices 
\bea
E_{\underline{m}}{}^{\underline{a}}&=&
\left(\begin{array}{cc}
	E_{4}{}^{4}&E_{4}{}^{\bar{a}}
	\\E_{\bar{m}}{}^{4}	&E_{\bar{m}}{}^{\bar{a}}	\end{array}	\right)~=~{\bf e}^{-1/4}
\left(\begin{array}{cc}1&-\tilde{B}^{\bar{n}}	e_{\bar{n}}{}^{\bar{a}}\\0&e_{\bar{m}}{}^{\bar{a}}	\end{array}	\right)
~~\label{Btilde}\\
&&\tilde{B}^{\bar{m}}=\frac{1}{2}\epsilon^{\bar{m}\bar{n}\bar{l}}
B_{\bar{n}\bar{l}}~~,~~{\bf e}={\rm det}~e_{\bar{m}}{}^{\bar{a}}~~~.\nn
\eea
The background gauge field in the tensor index is now
\bea
G_{MN}&=&G_{\underline{m}_1\underline{m}_2;\underline{n}_1\underline{n}_2}~=~
\frac{1}{2^2}E_{\underline{m}_1\underline{m}_2}{}^{\underline{a}_1\underline{a}_2}\hat{\eta}_{\underline{a}_1[\underline{b}_1}\hat{\eta}_{\underline{b}_2]\underline{a}_2}
E_{\underline{n}_1\underline{n}_2}{}^{\underline{b}_1\underline{b}_2}
\nn\\&=&\left(\begin{array}{cc}
	G_{\bar{m}\bar{n}}
	&G_{\bar{m};\bar{n}_1\bar{n}_2}\\
	G_{\bar{m}_1\bar{m}_2;\bar{n}}&
	G_{\bar{m}_1\bar{m}_2;\bar{n}_1\bar{n}_2}	\end{array}	\right)\nn\\
&=&{\bf  e}^{-1}\left(\begin{array}{cc}
	g_{\bar{m}\bar{n}}-\tilde{B}^{\bar{p}}g_{\bar{m}[\bar{p}}g_{\bar{n}]\bar{q}}\tilde{B}^{\bar{q}}
	&g_{\bar{m}[\bar{n}_1}g_{\bar{n}_2]\bar{l}}\tilde{B}^{\bar{l}}\\
	-\tilde{B}^{\bar{p}}g_{\bar{p}[\bar{m}_1}g_{\bar{m}_2]\bar{n}}&
	g_{\bar{m}_1[\bar{n}_1}g_{\bar{n}_2]\bar{m}_2}	\end{array}	\right)\label{BGtensor}
\eea
where metric of the stability group is denoted as $\hat{\eta}_{\underline{m}\underline{n}}$ 
to distinguish from 
$\eta_{MN}$.

The ${\cal T}$-string Lagrangian is obtained from the world-volume covariant 
${\cal A}$5-Lagrangian \bref{ALagwithSL6}
\bea
I&=&\displaystyle\int d^2\sigma~L\nn\\
L&=&\frac{\Phi}{2}\left(
-(F^{\hat{0}\underline{a}_1\underline{a}_2})^2+
(F^{\hat{5}\underline{a}_1\underline{a}_2})^2\right)
+\frac{1}{2}\Lambda_{\hat{0}\hat{5}}F^{\hat{0}\underline{a}_1\underline{a}_2}F^{\hat{5}}{}_{\underline{a}_1\underline{a}_2}\nn\\
&&
~~~~~~~~~~~~~~~~~~~~~
+\frac{1}{2}\Lambda_{\underline{a}\underline{b}}F^{\underline{a}\hat{0}\underline{c}}F^{\underline{b}}{}_{\hat{0}\underline{c}}
+\frac{1}{2}\Lambda_{\underline{a}\underline{b}}F^{\underline{a}\hat{5}\underline{c}}F^{\underline{b}}{}_{\hat{5}\underline{c}}\nn\\
&&+\frac{\epsilon_{\hat{0}\hat{5}\underline{a}_1\cdots \underline{a}_4}}{4}
\left(\tilde{\Lambda}_{\hat{0}}{}^{\hat{0}}F^{\hat{5}\underline{a}_1\underline{a}_2}F^{\hat{0}\underline{a}_3\underline{a}_4}
+\tilde{\Lambda}_{\hat{5}}{}^{\hat{0}}F^{\hat{5}\underline{a}_1\underline{a}_2}F^{\hat{5}\underline{a}_3\underline{a}_4} \nn\right.\\
&&
\left.~~~~~~~~~~~~~~~~~~~~~-\tilde{\Lambda}_{\hat{0}}{}^{\hat{5}}F^{\hat{0}\underline{a}_1\underline{a}_2}F^{\hat{0}\underline{a}_3\underline{a}_4}
-\tilde{\Lambda}_{\hat{5}}{}^{\hat{5}}F^{\hat{0}\underline{a}_1\underline{a}_2}F^{\hat{5}\underline{a}_3\underline{a}_4}
\right)
\eea
with $\eta^{\hat{0}\hat{0}} =-1$ and  $\eta^{\hat{5}\hat{5}} =1$.
Although currents are written as field strengths, there is no gauge symmetry of the coordinate $\delta X^{\underline{m}\underline{n}}$.
The ${\cal T}$-string Lagrangian in the SL(4) tensor coordinate is given by  
\bea
L&=&\phi \frac{1}{2^2}J_{\rm SD}{}^{\underline{a}_1\underline{a}_2}\hat{\eta}_{\underline{a}_1[\underline{b}_1}\hat{\eta}_{\underline{b}_2]\underline{a}_2}
J_{\overline{\rm SD}}{}^{\underline{b}_1\underline{b}_2}
+\bar{\phi}\frac{1}{2^2} J_{\overline{\rm SD}}{}^{\underline{a}_1\underline{a}_2}\hat{\eta}_{\underline{a}_1[\underline{b}_1}\hat{\eta}_{\underline{b}_2]\underline{a}_2}J_{\overline{\rm SD}}{}^{\underline{b}_1\underline{b}_2}\nn\\&&
+\tilde{\phi} \frac{1}{2^2}J_{\overline{\rm SD}}{}^{\underline{a}_1\underline{a}_2}{\epsilon}_{\underline{a}_1\cdots\underline{a}_4}J_{\overline{\rm SD}}{}^{\underline{a}_3\underline{a}_4}\nn\\
&=&\phi \frac{1}{2^2}\don\circ{J}_{\rm SD}{}^{\underline{m}_1\underline{m}_2}G_{{\underline{m}_1\underline{m}_2};{\underline{n}_1\underline{n}_2}}\don\circ{J}_{\overline{\rm SD}}{}^{\underline{n}_1\underline{n}_2}
+\bar{\phi} \frac{1}{2^2}\don\circ{J}_{\overline{\rm SD}}{}^{\underline{m}_1\underline{m}_2}G_{{\underline{m}_1\underline{m}_2};{\underline{n}_1\underline{n}_2}}\don\circ{J}_{\overline{\rm SD}}{}^{\underline{n}_1\underline{n}_2}\nn\\&&
+\tilde{\phi} \frac{1}{2^2}\don\circ{J}_{\overline{\rm SD}}{}^{\underline{m}_1\underline{m}_2}{\epsilon}_{\underline{m}_1\cdots\underline{m}_4}\don\circ{J}_{\overline{\rm SD}}{}^{\underline{m}_3\underline{m}_4}\label{LagSL4st}
~~~
\eea
with the background metric $G_{\underline{m}_1\underline{m}_2;\underline{n}_1\underline{n}_2}$
in \bref{BGtensor}.

The ${\cal T}$-string Lagrangian in the gauge $\Phi=g^2{\bf e}$ and
$\Lambda_{\hat{a}\hat{b}}=0=\tilde{\Lambda}_{\hat{a}}{}^{\hat{b}}$ as
\bea
L&=&-\frac{g^2{\bf e}}{2}\left(
(F^{\hat{0}\underline{a}_1\underline{a}_2})^2-
(F^{\hat{5}\underline{a}_1\underline{a}_2})^2\right)~~~.\label{TstLagfromA5}
\eea
The SL(4) covariant Lagrangian \bref{LagSL4st} in the gauge $\phi=\frac{1}{2g}$ and $\bar{\phi}=0=\tilde{\phi}$ 
is given as
\bea
L&=&\displaystyle\frac{1}{2g}~ J_{\rm SD}{}^A\hat{\eta}_{AB}J_{\overline{\rm SD}}{}^B\nn\\
&=&
\displaystyle\frac{1}{2^3g}~\varepsilon_+X^{\underline{m}_1\underline{m}_2}~ G_{\underline{m}_1\underline{m}_2;\underline{n}_1\underline{n}_2} ~\varepsilon_- X^{\underline{n}_1\underline{n}_2}\nn\\
&=&\displaystyle\frac{1}{2g}~(\varepsilon_+ x^{\bar{m}}~\varepsilon_+y^{\bar{m}_1\bar{m}_2})
\left(\begin{array}{cc}
	g_{\bar{m}\bar{n}}-\tilde{B}^{\bar{p}}g_{\bar{m}[\bar{p}}g_{\bar{n}]\bar{q}}\tilde{B}^{\bar{q}}
	&g_{\bar{m}[\bar{n}_1}g_{\bar{n}_2]\bar{l}}\tilde{B}^{\bar{l}}\\
	-\tilde{B}^{\bar{p}}g_{\bar{p}[\bar{m}_1}g_{\bar{m}_2]\bar{n}}&
	g_{\bar{m}_1[\bar{n}_1}g_{\bar{n}_2]\bar{m}_2}	\end{array}	\right)
\left(
\begin{array}{c}
	\varepsilon_- x^{\bar{m}} \\\varepsilon_- y^{\bar{m}_1\bar{m}_2}
\end{array}
\right)~~~.\nn\\
\label{SL4kineticxy}
\eea

\par
\vskip 6mm
\subsection{${S}$tring from ${\cal T}$-string }

We break SL(4) symmetry of ${\cal T}$-string into GL(3) for the usual string, where the reduction of the spacetime coordinate is performed as $X^{\underline{m}\underline{n}}=(X^{4\bar{m}},X^{\bar{m}\bar{n}})=(x^{\bar{m}},y^{\bar{m}\bar{n}})$ $\to$ $x^{\bar{m}}$.  
We repeat the same procedure of subsection \ref{Reduction2st}.
The SL(4) Lagrangian \bref{SL4kineticxy}
is rewritten analogously to \bref{kineticTst}
\bea
&&\displaystyle\frac{1}{2g}~ J_{\rm SD}{}^A\hat{\eta}_{AB}J_{\overline{\rm SD}}{}^B\nn\\
&&~~~=~\displaystyle\frac{1}{2g}\varepsilon_+x^{\bar{m}} g_{\bar{m}\bar{n}}\varepsilon_- x^{\bar{n}}\nn\\
&&~~~~~+\frac{1}{2^3g}(\varepsilon_+y^{\bar{m}_1\bar{m}_2} - \varepsilon_+x^{[\bar{m}_1}\tilde{B}^{\bar{m}_2}])g_{\bar{m}_1[\bar{n}_1}g_{\bar{n}_2]\bar{m}_2}
(\varepsilon_-y^{\bar{n}_1\bar{n}_2} +\tilde{B}^{[\bar{n}_1}\varepsilon_-x^{\bar{n}_2]})
\eea
By the dimensional reduction \bref{dimreduction} the Lagrangian \bref{SL4kineticxy} reduces to
the kinetic term of the string \bref{kinusualst}.

The total derivative term which is added to obtain the Wess-Zumino term 
\bref{totalderst} becomes
\bea
&&-\frac{1}{2^2g}(\varepsilon_+x^{\bar{m}_1}\varepsilon_- y^{\bar{m}_2\bar{m}_3}
-\varepsilon_-x^{\bar{m}_1}\varepsilon_+ y^{\bar{m}_2\bar{m}_3})
\epsilon_{\bar{m}_1\bar{m}_2\bar{m}_3}\nn\\&&~~~=~
-\frac{1}{2}\partial_\mu(\epsilon^{\mu\nu}x^{\bar{m}_1}\partial_\nu y^{\bar{m}_2\bar{m}_3}\epsilon_{\bar{m}_1\bar{m}_2\bar{m}_3})\nn\\
&&~~~=~
\frac{1}{2^2}\partial_\mu(\epsilon^{\mu\nu}X^{\underline{m}_1\underline{m}_2}\partial_\nu X^{\underline{m}_3\underline{m}_4}
\epsilon_{\underline{m}_1\cdots\underline{m}_4})
~~~.\label{totalderSL4st}
\eea
Adding this term to the SL(4) Lagrangian \bref{SL4kineticxy}
\bea
&&\frac{1}{2g} J_{\rm SD}{}^A\hat{\eta}_{AB}J_{\overline{\rm SD}}{}^B
+\frac{1}{2^2}\partial_\mu(\epsilon^{\mu\nu}
X^{\underline{m}_1\underline{m}_2}\partial_\nu X^{\underline{m}_3\underline{m}_4}\epsilon_{\underline{m}_1\cdots\underline{m}_4})\label{SL4stringLagWZ}\\
&&~~~=
\frac{1}{2g}
\left\{
\varepsilon_+x^{\bar{m}} g_{\bar{m}\bar{n}} \varepsilon_- x^{\bar{n}}\right.\nn\\
&&~~~~~\left.+\frac{1}{2^2}
(\varepsilon_+ y^{\bar{m}_1\bar{m}_2} -\varepsilon_+ x^{[\bar{m}_1} \tilde{B}^{\bar{m}_2]}+\varepsilon_+ x^{\bar{l}_3}\epsilon_{\bar{l}_1\bar{l}_2\bar{l}_3} g^{\bar{l}_1\bar{m}_1} g^{\bar{l}_2\bar{m}_2}) 
g_{\bar{m}_1[\bar{n}_1} g_{\bar{n}_2]\bar{m}_2} \right.\nn\\
&&~~~~~~\left.~\times(\varepsilon_- y^{\bar{n}_1\bar{n}_2} +\tilde{B}^{[\bar{n}_1}\varepsilon_- x^{\bar{n}_2]} 
-g^{\bar{n}_1\bar{k}_1} g^{\bar{n}_2\bar{k}_2}\epsilon_{\bar{k}_1\bar{k}_2\bar{k}_3} \varepsilon_- x^{\bar{k}_3}) 
\right.\nn\\
&&~~~~~~~\left.+\varepsilon_+x^{\bar{m}} g_{\bar{m}\bar{n}} \varepsilon_- x^{\bar{n}}+2\varepsilon_+x^{\bar{m}} B_{\bar{m}\bar{n}} \varepsilon_- x^{\bar{n}}\right\}~~~\nn
\eea
After the dimensional reduction \bref{dimreduction}, 
the Lagrangian with the total derivative term reduces into the usual string Lagrangian  with the Wess-Zumino term \bref{GLDstLag},
\bea
L_0+L_{\rm WZ}&=&\frac{1}{g}~\varepsilon_+x^{\bar{m}} (g_{\bar{m}\bar{n}}+B_{\bar{m}\bar{n}}) \varepsilon_- x^{\bar{n}}~~~.\nn
\eea

\par
\vskip 6mm
\section{Lagrangians of M2-brane via ${\cal M}$5-brane
	from ${\cal A}$5-brane}\label{section:6}

\par
\vskip 6mm
\subsection{${\cal M}$5-brane from ${\cal A}$5-brane  }\label{section:6-1}

The GL(4) ${\cal M}$5-brane from ${\cal A}$5-brane is 
described by  the GL(4) vector 
coordinate $X^{5\underline{m}}=x^{\underline{m}}$ \cite{Linch:2015fya} as listed in \bref{rep}.
The SL(6) rank-two tensor coordinate is decomposed as SL(6) $\to$ SL(5) $\to$ GL(4)  as 
$X^{\hat{m}\hat{n}}=(X^{0n}=Y^n,~X^{mn})$ $\to$
$X^{{m}{n}}=(X^{5\underline{n}}=x^{\underline{m}},~
X^{\underline{m}\underline{n}}=y^{\underline{m}\underline{n}})$ 
and $Y^m=(Y^5=Y,~Y^{\underline{m}})$.
The 6-dimensional world-volume derivative is reduced into the 5-brane  world-sheet derivatives as 
$\partial^{\hat{m}}=(\partial^0=\partial_\tau,~\partial^5=0,~\partial^{\underline{m}}=\partial_\sigma{}^{\underline{m}})$. 
The SL(6) field strength for the ${\cal M}$5-brane has the following components
\bea
{\renewcommand{\arraystretch}{1.2}
\left\{
\begin{array}{ccl}
\don\circ{F}{}^{05\underline{m}}&=&\partial_\tau x^{\underline{m}}+\partial^{\underline{m}}Y \\
\don\circ{F}{}^{5\underline{mn}}&=&-\partial^{[\underline{m}}x^{\underline{n}]}	\\
\don\circ{F}{}^{0\underline{mn}}&=&\partial_{\tau}y^{\underline{m}\underline{n}} -\partial^{[\underline{m}}Y^{\underline{n}]}\\
\don\circ{F}{}^{\underline{mnl}}&=&\frac{1}{2}\partial^{[\underline{m}}y^{\underline{n}\underline{l}]}
\end{array}\right.}
\label{SL5M5}~~~
\eea
where the auxiliary coordinates $y^{\underline{m}\underline{n}}$ and $Y^{m}$ are preserved to begin with the SL(5) $A$-symmetric ${\cal M}$-theory Lagrangian
\cite{Siegel:2020qef}.

The SL(6) vielbein for the ${\cal M}$5-brane with SL(5) $A$-symmetry is given by
\bea
E_{\hat{m}}{}^{\hat{a}}&=&
{\renewcommand{\arraystretch}{1.8}
	\left(\begin{array}{cc}
		E_0{}^{\hat{0}}&E_{0}{}^{{a}}\\
E_m{}^{\hat{0}}&E_m{}^a		
	\end{array}\right)=
	\left(\begin{array}{cc}
		\displaystyle\frac{1}{g}&0\\
	-\displaystyle\frac{s_m}{g}&g^{1/5}E_m{}^a	
	\end{array}\right)}
%
\label{M5A5vielbein}~~~.
\eea
It is stressed that the world-volume vielbein fields $g$,  $s_m$ and the spacetime vielbein $E_m{}^a$ cannot be in block diagonal form unlike ${\cal T}$-string case
\bref{TstinA5}.
The selfdual and anti-selfdual currents in curved background given by 
 \bref{SL5SDcurvedcurrent} based on \bref{SL5SDflatcurrent} 
are the following combination of the SL(6) field strengths in \bref{SL5M5} with \bref{SL6currents} as
\bea
{F}_{{\rm SD}/\overline{\rm SD}}{}^{{a}_1{a}_2}&=&g\left( F^{\hat{0}{a}_1{a}_2}
\pm \frac{1}{3!} \epsilon^{\hat{0}{a}_1{a}_2}{}_{{a}_3{a}_4a_5} 
F^{{a}_3{a}_4a_5}\right)~~~.
\eea
The GL(4) covariant selfdual and anti-selfdual currents in flat space  are derived from the ones of SL(5)  \bref{SL5SDflatcurrent} given in \cite{Hatsuda:2023dwx} as
\bea
\left\{\begin{array}{lcl}
	\don\circ{F}_{{\rm SD}/\overline{\rm SD}}{}^{\underline{m}}&=&
	{F}_{\tau}{}^{\underline{m}}
	-\frac{1}{2}\epsilon^{\underline{m}\underline{n}_1\underline{n}_2\underline{n}_3}s_{\underline{n}_1}{F}_{\sigma;\underline{n}_2\underline{n}_3}\pm g \hat{\eta}^{\underline{m}\underline{n}}{F}_{\sigma;\underline{n}}\\
&=&\varepsilon_\tau{}x^{\underline{m}}\pm g \hat{\eta}^{\underline{m}\underline{n}}
	(\varepsilon_{\sigma} y)_{\underline{n}}
	\\
	\don\circ{F}_{{\rm SD}/\overline{\rm SD}}{}^{\underline{m}_1\underline{m}_2}
	&=&	{F}_{\tau}{}^{\underline{m}_1\underline{m}_2}
	+\epsilon^{\underline{m}_1\cdots\underline{m}_4}s_{\underline{m}_3}{F}_{\sigma;\underline{m}_4}
	-\frac{1}{2}\epsilon^{\underline{m}_1\cdots\underline{m}_4}s_5F_{\sigma;\underline{m}_3\underline{m}_4}
	\pm g \hat{\eta}^{\underline{m}_1\underline{n}_1}
	\hat{\eta}^{\underline{m}_2\underline{n}_2}
	{F}_{\sigma;\underline{n}_1\underline{n}_2}\\
&=&\varepsilon_\tau{}y^{\underline{m}_1\underline{m}_2}\pm g \hat{\eta}^{\underline{m}_1\underline{n}_1}\hat{\eta}^{\underline{m}_2\underline{n}_2}
	(\varepsilon_{\sigma}x)_{\underline{n}_1\underline{n}_2}
\end{array}\right.\label{GL4SDASDcurrents}
\eea
where $\hat{\eta}^{mn}$ becomes $G^{mn}$ in a curved background.  
The brane world-volume derivatives are given as a generalization of the world-sheet zweibein dependence in
\bref{zweibein} as
\bea
&&\left\{\begin{array}{lcl}
\varepsilon_\tau{}x^{\underline{m}}&\equiv&
	{F}_{\tau}{}^{\underline{m}}
	-\frac{1}{2}\epsilon^{\underline{m}\underline{n}_1\underline{n}_2\underline{n}_3}s_{\underline{n}_1}{F}_{\sigma;\underline{n}_2\underline{n}_3}\\
	&=&	\dot{x}^{\underline{m}}+\partial^{\underline{m}}Y+s_{\underline{n}}
	\partial^{[\underline{m}}x^{\underline{n}]}\\
\varepsilon_\tau{}y^{\underline{m}_1\underline{m}_2}&\equiv&
	{F}_{\tau}{}^{\underline{m}_1\underline{m}_2}
	+\epsilon^{\underline{m}_1\cdots\underline{m}_4}s_{\underline{m}_3}{F}_{\sigma;\underline{m}_4}\\
	&=&\dot{y}^{\underline{m}_1\underline{m}_2}-\partial^{[\underline{m}_1}Y^{\underline{m}_2]}+\frac{1}{2}s_{\underline{m}_3}\partial^{[\underline{m}_1}y^{\underline{m}_2\underline{m}_3]}{+s_5\partial^{[\underline{m}_1}x^{\underline{m}_2]}}
\end{array}\right.~\nn\\
&&\left\{\begin{array}{lcl}
	(\varepsilon_{\sigma}y)_{\underline{m}}&\equiv& {F}_{\sigma;\underline{m}}\\
	&=&\frac{1}{2}\epsilon_{\underline{m}\underline{n}_1\underline{n}_2\underline{n}_3}\partial^{\underline{n}_1}y^{\underline{n}_2\underline{n}_3}\\
(\varepsilon_{\sigma}x)_{\underline{m}_1\underline{m}_2}&\equiv&
	{F}_{\sigma;\underline{m}_1\underline{m}_2}
	\\&=&
	-\epsilon_{\underline{m}_1\cdots \underline{m}_4}\partial^{\underline{m}_3}x^{\underline{m}_4}
\end{array}\right.~~~.
\eea

The 11-dimensional supergravity background includes
the gravitational metric $g_{mn}$ and the three form gauge field $C_{mnl}$.
We focus on the 4-dimensional subspace of the 11-dimensional space, where the background fields are $g_{\underline{m}\underline{n}}$ and $C_{\underline{m}\underline{n}\underline{l}}$
whose number of degrees of freedom is $10+4=14$.
The dimension of the coset SL(5)/SO(5) is also 
$24-10=14$. 
The vector vielbein  $E_{m}{}^a\in$ SL(5)/SO(5) 
with GL(4) indices where $m=(5,\underline{m})$ and $\underline{m}=1,\cdots,4$ 
is given by \cite{Hatsuda:2012vm} 
\bea
E_m{}^a
&=&
\left(\begin{array}{cc}
	E_{5}{}^{5}&E_{5}{}^{\underline{a}}
	\\E_{\underline{m}}{}^{5}	&E_{\underline{m}}{}^{\underline{a}}	\end{array}	\right)~=~
\left(\begin{array}{cc}{\bf e}^{3/5}&{\bf e}^{-2/5}\tilde{C}^{\underline{n}}	e_{\underline{n}}{}^{\underline{a}}\\0&{\bf e}^{-2/5}e_{\underline{m}}{}^{\underline{a}}	\end{array}	\right)\nn\\
&&\tilde{C}^{\bar{m}}=\frac{1}{3!}\epsilon^{\underline{m}\underline{m}_1\underline{m}_2\underline{m}_3}C_{\underline{m}_1\underline{m}_2\underline{m}_3}~~,~~
{\bf e}=\det e_{\underline{m}}{}^{\underline{a}}~~~
\label{Ctilde}
\eea
with $\det E_{m}{}^a=1= \epsilon^{m_1\cdots m_5}E_{m_1}{}^{a_1}E_{m_2}{}^{a_2}E_{m_3}{}^{a_3}E_{m_4}{}^{a_4}E_{m_5}{}^{a_5}= \epsilon^{a_1\cdots a_5}$.
The tensor vielbein is the product of the vector vielbein \bref{Ctilde} as
\bea
E_{{m}_1{m}_2}{}^{{a}_1{a}_2}
&=&
E_{[{m}_1}{}^{{a}_1}
E_{{m}_2]}{}^{{a}_2}
\nn\\
&=&
\left(\begin{array}{cc}
	E_{5\underline{m}}{}^{5\underline{a}}&E_{5\underline{m}}{}^{\underline{a}_1\underline{a}_2}
	\\
	E_{\underline{m}_1\underline{m}_2}{}^{5\underline{a}}	&E_{\underline{m}_1\underline{m}_2}{}^{\underline{a}_1\underline{a}_2}	\end{array}	\right)~=~
\left(\begin{array}{cc}{\bf e}^{1/5}e_{\underline{m}}{}^{\underline{a}}
	&-{\bf e}^{-4/5}\tilde{C}^{\underline{n}}	e_{\underline{m}}{}^{[\underline{a}_1}
	e_{\underline{n}}{}^{\underline{a}_2]}
	\\0&{\bf e}^{-4/5}e_{\underline{m}_1}{}^{[\underline{a}_1}e_{\underline{m}_2}{}^{\underline{a}_2]}	\end{array}	\right)~~~.\nn\\
\eea
The background gauge field in tensor index is now
\bea
G_{MN}&=&G_{{m}_1{m}_2;{n}_1{n}_2}~=~
\frac{1}{2^2}E_{{m}_1{m}_2}{}^{{a}_1{a}_2}\hat{\eta}_{{a}_1[{b}_1}\hat{\eta}_{{b}_2]{a}_2}
E_{{n}_1{n}_2}{}^{{b}_1{b}_2}
\nn\\&=&\left(\begin{array}{cc}
	G_{\underline{m}\underline{n}}
	&G_{\underline{m};\underline{n}_1\underline{n}_2}\\
	G_{\underline{m}_1\underline{m}_2;\underline{n}}&
	G_{\underline{m}_1\underline{m}_2;\underline{n}_1\underline{n}_2}	\end{array}	\right)\nn\\
&=&{\bf e}^{-8/5}\left(\begin{array}{cc}
	{\bf e}^{2}g_{\underline{m}\underline{n}}-\tilde{C}^{\underline{p}}g_{\underline{m}[\underline{p}}g_{\underline{n}]\underline{q}}\tilde{C}^{\underline{q}}
	&\tilde{C}^{\underline{l}}g_{\underline{l}[\underline{n}_1}g_{\underline{n}_2]\underline{m}}\\
	g_{\underline{p}[\underline{m}_1}g_{\underline{m}_2]\underline{n}}	\tilde{C}^{\underline{p}}&
	g_{\underline{m}_1[\underline{n}_1}g_{\underline{n}_2]\underline{m}_2}	\end{array}	\right)\nn\\
&=&
{\bf e}^{2/5}\left(\begin{array}{cc}
	g_{\underline{m}\underline{n}}&0\\0&0
\end{array}	\right)
+{\bf e}^{-8/5}\left(\begin{array}{cc}
	-\tilde{C}^{\underline{p}}g_{\underline{m}[\underline{p}}g_{\underline{n}]\underline{q}}\tilde{C}^{\underline{q}}
	&\tilde{C}^{\underline{l}}g_{\underline{l}[\underline{n}_1}g_{\underline{n}_2]\underline{m}}\\
	g_{\underline{p}[\underline{m}_1}g_{\underline{m}_2]\underline{n}}	\tilde{C}^{\underline{p}}&
	g_{\underline{m}_1[\underline{n}_1}g_{\underline{n}_2]\underline{m}_2}	\end{array}	\right)
\label{BGtensorC}~~~.
\eea
Inverse of these background gauge fields are given by \cite{Hatsuda:2012vm} 
\bea
E_a{}^m
&=&
\left(\begin{array}{cc}
	E_{5}{}^{5}&E_{5}{}^{\underline{m}}
	\\E_{\underline{a}}{}^{5}	&E_{\underline{a}}{}^{\underline{m}}	\end{array}	\right)~=~
\left(\begin{array}{cc}{\bf e}^{-3/5}&{\bf e}^{-3/5}\tilde{C}^{\underline{m}}	\\0&{\bf e}^{2/5}e_{\underline{a}}{}^{\underline{m}}	\end{array}	\right)\\
E_{{a}_1{a}_2}{}^{{m}_1{m}_2}
&=&
E_{[{a}_1}{}^{{m}_1}
E_{{a}_2]}{}^{{m}_2}\nn\\
&=&
\left(\begin{array}{cc}
	E_{5\underline{a}}{}^{5\underline{m}}&E_{5\underline{a}}{}^{\underline{m}_1\underline{m}_2}
	\\
	E_{\underline{a}_1\underline{a}_2}{}^{5\underline{m}}	&E_{\underline{a}_1\underline{a}_2}{}^{\underline{m}_1\underline{m}_2}	\end{array}	\right)~=~
\left(\begin{array}{cc}e_{\underline{a}}{}^{\underline{m}}
	&-\tilde{C}^{[\underline{m}_1}	e_{\underline{a}}{}^{[\underline{m}_2]}
	\\0&{\bf e}e_{\underline{a}_1}{}^{[\underline{m}_1}e_{\underline{a}_2}{}^{\underline{m}_2]}	\end{array}	\right)~~\\
G^{MN}&=&G^{{m}_1{m}_2;{n}_1{n}_2}~=~
\frac{1}{2^2}E_{{a}_1{a}_2}{}^{{m}_1{m}_2}\hat{\eta}^{{a}_1[{b}_1}\hat{\eta}^{{b}_2]{a}_2}
E_{{b}_1{b}_2}{}^{{n}_1{n}_2}
\nn\\&=&\left(\begin{array}{cc}
	G^{\underline{m}\underline{n}}
	&G^{\underline{m};\underline{n}_1\underline{n}_2}\\
	G^{\underline{m}_1\underline{m}_2;\underline{n}}&
	G^{\underline{m}_1\underline{m}_2;\underline{n}_1\underline{n}_2}	\end{array}	\right)\nn\\
&=&{\bf e}^{2/5}\left(\begin{array}{cc}
	g^{\underline{m}\underline{n}}
	&g^{\underline{m}[\underline{n}_1} \tilde{C}^{\underline{n}_2]}\\
	-\tilde{C}^{[\underline{m}_1}g^{\underline{m}_2]\underline{n}}	&
	{\bf e}^{2}	g^{\underline{m}_1[\underline{n}_1}g^{\underline{n}_2]\underline{m}_2}	
	+\tilde{C}^{[\underline{m}_1}g^{\underline{m}_2][\underline{n}_1}\tilde{C}^{\underline{n}_2]}
\end{array}	\right)\nn\\
&=&{\bf e}^{8/5}\left(\begin{array}{cc}
	0	&0\\0&	g^{\underline{m}_1[\underline{n}_1}g^{\underline{n}_2]\underline{m}_2}	
\end{array}	\right)+
{\bf e}^{-2/5}\left(\begin{array}{cc}
	g^{\underline{m}\underline{n}}
	&g^{\underline{m}[\underline{n}_1} \tilde{C}^{\underline{n}_2]}\\
	-\tilde{C}^{[\underline{m}_1}g^{\underline{m}_2]\underline{n}}	&
	\tilde{C}^{[\underline{m}_1}g^{\underline{m}_2][\underline{n}_1}\tilde{C}^{\underline{n}_2]}
\end{array}	\right)\label{BGtensorCinverse}~~~.
\eea

The ${\cal M}$5-brane Lagrangian is given 
by the SL(5) covariant Lagrangian \bref{ALagwithSL5} 
with replacing GL(4) indices as
\bea
L&=&
\frac{1}{2}\phi 
\left(
F_{\rm SD}{}^{\underline{a}}F_{\overline{\rm SD}\underline{a}}
+
\frac{1}{2}
F_{\rm SD}{}^{\underline{a}\underline{b}}
F_{\overline{\rm SD}\underline{a}\underline{b}}
\right)
+\frac{1}{2}\bar{\phi} 
\left(
(F_{\overline{\rm SD}}{}^{\underline{a}})^2
+
\frac{1}{2}(F_{\overline{\rm SD}}{}^{\underline{a}\underline{b}})^2
\right)\nn\\&&
+\frac{1}{2}\lambda F_{\overline{\rm SD}}{}^{\underline{c}}
F_{\overline{\rm SD}}{}^{}{}_{\underline{c}}
+\lambda_{\underline{a}}
F_{\overline{\rm SD}}{}^{\underline{a}\underline{c}}
F_{\overline{\rm SD}}{}^{}{}_{\underline{c}}
+\frac{1}{2}\lambda_{\underline{a}\underline{b}}
\left(
F_{\overline{\rm SD}}{}^{\underline{a}}
F_{\overline{\rm SD}}{}^{\underline{b}}{}
+
F_{\overline{\rm SD}}{}^{\underline{a}\underline{c}}
F_{\overline{\rm SD}}{}^{\underline{b}}{}_{\underline{c}}
\right)
\nn\\&&
-\epsilon_{\underline{a}_1\cdots \underline{a}_4}
\left( \frac{1}{8}\lambda^{5}
F_{\overline{\rm SD}}{}^{\underline{a}_1\underline{a}_2}
F_{\overline{\rm SD}}{}^{\underline{a}_3\underline{a}_4}
-\frac{1}{2}
\lambda^{\underline{a}_1}
F_{\overline{\rm SD}}{}^{\underline{a}_2\underline{a}_3}
F_{\overline{\rm SD}}{}^{\underline{a}_4}
\right)
\label{MLagwithGL4}~~~.
\eea
The Lagrangian in terms of the curved currents 
$F_{\rm SD/\overline{\rm SD}}^{ab}$ 
is simpler than the one in terms of the flat currents $\don\circ{F}_{\rm SD/\overline{\rm SD}}{}^{mn}$.
The concrete expression of the Lagrangian of the ${\cal M }$5-brane in a curved background \bref{MLagwithGL4} is
given as follows.
We begin by the SL(5) covariant Lagrangian \bref{ALagwithSL5} in the gauge $\phi=\frac{1}{2g}$ and $\bar{\phi}=0=\lambda$'s
\bea
&&\displaystyle\frac{1}{2g}~ 
{F}_{\rm SD}{}^{A}\hat{\eta}_{AB}F_{\overline{\rm SD}}{}^B\nn\\
&&~=~\displaystyle\frac{1}{8g}~ 
\don\circ{F}_{\rm SD}{}^{m_1m_2}G_{{m}_1{m}_2;{n}_1{n}_2}\don\circ{F}_{\overline{\rm SD}}{}^{n_1n_2}\nn\\
&&~=~\displaystyle\frac{1}{8g}
\varepsilon_\tau{}X^{m_1m_2}G_{{m}_1{m}_2;{n}_1{n}_2}\varepsilon_\tau{}X^{n_1n_2}
-\displaystyle\frac{g}{8}
F_\sigma{}_{m_1m_2}G^{{m}_1{m}_2;{n}_1{n}_2}F_\sigma{}_{n_1n_2}
\nn\\
&&~=~\displaystyle\frac{1}{2g}~
(\varepsilon_\tau x^{\underline{m}}~\varepsilon_\tau y^{\underline{m}_1\underline{m}_2})
\left[
{\bf e}^{2/5}\left(\begin{array}{cc}
	g_{\underline{m}\underline{n}}&0\\0&0
\end{array}	\right)\right.\nn\\
&&~~~~~~~~~~~~~~~~~~~~~~~~~~~~~~~~~~~~\left.
+{\bf e}^{-8/5}\left(\begin{array}{cc}
	-\tilde{C}^{\underline{p}}g_{\underline{m}[\underline{p}}g_{\underline{n}]\underline{q}}\tilde{C}^{\underline{q}}
	&\tilde{C}^{\underline{l}}g_{\underline{l}[\underline{n}_1}g_{\underline{n}_2]\underline{m}}\\
	g_{\underline{p}[\underline{m}_1}g_{\underline{m}_2]\underline{n}}	\tilde{C}^{\underline{p}}&
	g_{\underline{m}_1[\underline{n}_1}g_{\underline{n}_2]\underline{m}_2}	\end{array}	\right)
\right]
\left(
\begin{array}{c}
	\varepsilon_\tau x^{\underline{n}} 
	\\\varepsilon_\tau y^{\underline{n}_1\underline{n}_2}
\end{array}
\right)\nn\\
&&~~~-\displaystyle\frac{g}{2}~
\left((\varepsilon_{\sigma}y)_{\underline{m}}~
(\varepsilon_{\sigma}x)_{\underline{m}_1\underline{m}_2}\right)
\left[{\bf e}^{8/5}\left(\begin{array}{cc}
	0	&0\\0&	g^{\underline{m}_1[\underline{n}_1}g^{\underline{n}_2]\underline{m}_2}	
\end{array}	\right)\right.\nn\\&&~~~~~~~~~~~~~~~~~~~~~~~~~~~~~~~~~~~~\left.
+
{\bf e}^{-2/5}\left(\begin{array}{cc}
	g^{\underline{m}\underline{n}}
	&g^{\underline{m}[\underline{n}_1} \tilde{C}^{\underline{n}_2]}\\
	-\tilde{C}^{[\underline{m}_1}g^{\underline{m}_2]\underline{n}}	&
	\tilde{C}^{[\underline{m}_1}g^{\underline{m}_2][\underline{n}_1}\tilde{C}^{\underline{n}_2]}
\end{array}	\right)
\right]
\left(
\begin{array}{c}
	(\varepsilon_{\sigma}y)_{\underline{n}}\\
	(\varepsilon_{\sigma}x)_{\underline{n}_1\underline{n}_2}
\end{array}
\right)\nn\\
&&~=~L_{0}+L_y\label{L0Ly}
\eea
with
\bea
L_{0}&=&
\displaystyle\frac{{\bf e}^{2/5}}{2g}\varepsilon_\tau x^{\underline{m}}
g_{\underline{m}\underline{n}}\varepsilon_\tau x^{\underline{n}}
-\displaystyle\frac{g{\bf e}^{8/5}}{8}
(\varepsilon_{\sigma}x)_{\underline{m}_1\underline{m}_2}
g^{\underline{m}_1[\underline{n}_1}g^{\underline{n}_2]\underline{m}_2}	
(\varepsilon_{\sigma}x)_{\underline{n}_1\underline{n}_2}
\\
L_y&=&\displaystyle\frac{{\bf e}^{-8/5}}{8g}
\left(\varepsilon_\tau y^{\underline{m}_1\underline{m}_2}
-\varepsilon_\tau x^{[\underline{m}_1}\tilde{C}^{\underline{m}_2]}
\right)g_{\underline{m}_1[\underline{n}_1}g_{\underline{n}_2]\underline{m}_2}
\left(\varepsilon_\tau y^{\underline{n}_1\underline{n}_2}
+\tilde{C}^{[\underline{n}_1}\varepsilon_\tau x^{\underline{n}_2]}
\right)
\nn\\
&&-\displaystyle\frac{g{\bf e}^{-2/5}}{2}
\left((\varepsilon_\sigma y)_{\underline{m}}
+(\varepsilon_\sigma x)_{\underline{m}\underline{l}}\tilde{C}^{\underline{l}}
\right)g^{\underline{m}\underline{n}}
\left((\varepsilon_\sigma y)_{\underline{n}}
-\tilde{C}^{\underline{l}}(\varepsilon_\sigma x)_{\underline{l}\underline{n}}
\right)~~~.
\nn
\eea
In the gauge $g={\bf e}^{-3/5}$ 
Lagrangians take  simple form as
\bea
L_0&=&-\displaystyle\frac{{\bf e}}{2}
\left[F^{05\underline{a}}
\hat{\eta}_{\underline{a}\underline{b}}F^{05\underline{b}}
-\displaystyle\frac{1}{4}
F^{5\underline{a}_1\underline{a}_2}
\hat{\eta}_{\underline{a}_1[\underline{b}_1}\hat{\eta}_{\underline{b}_2]\underline{a}_2}	F^{5\underline{b}_1\underline{b}_2}\right]
\label{M5L0}
\\
L_y&=&-\displaystyle\frac{1}{2{\bf e}}~\left[\frac{1}{2}
F^{0\underline{a}_1\underline{a}_2}
\hat{\eta}_{\underline{a}_1\underline{b}_1}
\hat{\eta}_{\underline{b}_2\underline{a}_2}
F^{0\underline{b}_1\underline{b}_2}
-\displaystyle\frac{1}{6}~F^{\underline{a}_1\underline{a}_2\underline{a}_3}
\hat{\eta}_{\underline{a}_1\underline{b}_1}
\hat{\eta}_{\underline{a}_2\underline{b}_2}
\hat{\eta}_{\underline{a}_3\underline{b}_3}
F^{\underline{b}_1\underline{b}_2\underline{b}_3}\right]\nn
\eea
with  
$\tilde{m}=(0,\underline{m})$. 

The SL(5) U-duality symmetry of the Lagrangian \bref{M5L0} is broken to GL(4) symmetry by the dimensional reduction similarly to \bref{dimreduction}. 
Then the kinetic term of  
the new perturbative Lagrangian for a ${\cal M}$5-brane in the 4-dimensions is given by
\bea
L_{0}&=&\frac{{\bf e}}{2}\left[
(\dot{x}^{\underline{m}}+\partial^{\underline{m}}Y+s_{\underline{l}}\partial^{[\underline{m}}x^{\underline{l}]})
g_{\underline{m}\underline{n}}
(\dot{x}^{\underline{n}}+\partial^{\underline{n}}Y+s_{\underline{k}}\partial^{[\underline{n}}x^{\underline{k}]})
-\frac{1}{4}
\partial^{[\underline{m}_1}x^{\underline{m}_2]}
g_{\underline{m}_1[\underline{n}_1}
g_{\underline{n}_2]\underline{m}_2}
\partial^{[\underline{n}_1}x^{\underline{n}_2]}\right]
.\nn\\
\eea

The total derivative terms to obtain the Wess-Zumino term for the ${\cal M}$5-brane are given analogously to the string case
\bref{totalderst} with the gauge $\partial^{\underline{m}}s_5=0$ as 
\bea
&&\varepsilon_\tau x^{\underline{m}}(\varepsilon_\sigma y)_{\underline{m}}
-\frac{1}{2}(\varepsilon_\sigma x)_{\underline{m}_1\underline{m}_2}\varepsilon_\tau y^{\underline{m}_1\underline{m}_2}
\label{totalderM5}\\
&&~=~\frac{1}{2}\epsilon_{\underline{m}_1\cdots \underline{m}_4}\Bigl\{
\partial_\tau(x^{\underline{m}_1}\partial^{\underline{m}_2}y^{\underline{m}_3\underline{m}_4})
\nn\\&&~~~~
+\partial^{\underline{m}_1}(
x^{\underline{m}_2}\dot{y}^{\underline{m}_3\underline{m}_4}
+Y\partial^{\underline{m}_2}y^{\underline{m}_3\underline{m}_4}
-2x^{\underline{m}_2}\partial^{\underline{m}_3}Y^{\underline{m}_4}
-2s_5 x^{\underline{m}_2}\partial^{\underline{m}_3}x^{\underline{m}_4})
\Bigr\}
~~~\nn
\eea
where the $s_{\underline{n}}$ dependent terms are cancelled out because of the totally antisymmetricity of 5 indices
\bea
&&\epsilon_{\underline{m}_1\cdots \underline{m}_4}s_{\underline{n}}
\Bigl((\partial^{[\underline{m}_1}x^{\underline{n}]})
\partial^{\underline{m}_2}y^{\underline{m}_3\underline{m}_4}
+\frac{1}{2}(\partial^{\underline{m}_4}x^{\underline{m}_1})
\partial^{[\underline{n}}
y^{\underline{m}_2\underline{m}_3]}    \Bigr)\nn\\
&&~~=
\epsilon_{\underline{m}_1\cdots \underline{m}_4}s_{\underline{n}}\frac{1}{4!}
\partial^{[\underline{m}_1}x^{\underline{n}}
\partial^{\underline{m}_2}y^{\underline{m}_3\underline{m}_4]}
=0  ~~~.
\eea
Adding the total derivative term \bref{totalderM5}
to the ${\cal M}$5-brane Lagrangian 
\bref{MLagwithGL4} in gauge $\phi=\frac{1}{2g}$, $g=2{\bf e}^{-3/5}$, $\bar{\phi}=0=\lambda$'s the Lagrangian for the ${\cal M}5$-brane becomes 
\bea
&&\displaystyle\frac{1}{2g}~ 
{F}_{\rm SD}{}^{A}\hat{\eta}_{AB}F_{\overline{\rm SD}}{}^B
-\varepsilon_\tau x^{\underline{m}}(\varepsilon_\sigma y)_{\underline{m}}
+\frac{1}{2}(\varepsilon_\sigma x)_{\underline{m}_1\underline{m}_2}\varepsilon_\tau y^{\underline{m}_1\underline{m}_2}
\nn\\
&&~~~~~~~~=L_0+L_y+L_{\rm WZ}\nn\eea
\bea
L_0&=&\displaystyle\frac{{\bf e}}{4}\left[\varepsilon_\tau x^{\underline{m}}
g_{\underline{m}\underline{n}}\varepsilon_\tau x^{\underline{n}}
-\displaystyle\frac{1}{4}
(\varepsilon_{\sigma}x)_{\underline{m}_1\underline{m}_2}
g^{\underline{m}_1[\underline{n}_1}g^{\underline{n}_2]\underline{m}_2}	
(\varepsilon_{\sigma}x)_{\underline{n}_1\underline{n}_2}\right]
\nn\\
L_y&=&\displaystyle\frac{1}{4{\bf e}}
\left[
\frac{1}{4}\left(\varepsilon_\tau y^{\underline{m}_1\underline{m}_2}
-\varepsilon_\tau x^{[\underline{m}_1}\tilde{C}^{\underline{m}_2]}
+{\bf e} (\varepsilon_\sigma x)_{\underline{l}_1\underline{l}_2}g^{\underline{l}_1\underline{m}_1}g^{\underline{l}_2\underline{m}_2}
\right)g_{\underline{m}_1[\underline{n}_1}g_{\underline{n}_2]\underline{m}_2}\right.\nn\\
&&~~~~\left.
\times\left(\varepsilon_\tau y^{\underline{n}_1\underline{n}_2}
+\tilde{C}^{[\underline{n}_1}\varepsilon_\tau x^{\underline{n}_2]}+{\bf e}g^{\underline{n}_1\underline{k}_1}g^{\underline{n}_2\underline{k}_2} (\varepsilon_\sigma x)_{\underline{k}_1\underline{k}_2}
\right)\right.
\nn\\
&&\left.
-\left((\varepsilon_\sigma y)_{\underline{m}}
+(\varepsilon_\sigma x)_{\underline{m}\underline{l}}\tilde{C}^{\underline{l}}
+{\bf e}\varepsilon_\tau x^{\underline{l}}g_{\underline{l}\underline{m}}
\right)
g^{\underline{m}\underline{n}}
\left((\varepsilon_\sigma y)_{\underline{n}}
-\tilde{C}^{\underline{k}}(\varepsilon_\sigma  x)_{\underline{k}\underline{n}}+{\bf e}g_{\underline{n}\underline{k}}\varepsilon_\tau x^{\underline{k}}\right)\right]\nn\\
L_{\rm WZ}&=&
 \varepsilon_\tau x^{\underline{m}_1}\tilde{C}^{\underline{m}_2}
(\varepsilon_\sigma x)_{\underline{m}_1\underline{m}_2}
~~~.
\eea
Dimensional reduction $L_y\to 0$ gives the ${\cal M}$5-brane Lagrangian with the Wass-Zumino term.

The obtained new ${\cal M}$5-brane Lagrangian in the supergravity background \bref{stringL} is
\bea
L_{{\cal M}5}&=&L_0+L_{\rm WZ}\nn\\
L_0&=&\displaystyle\frac{{\bf e}}{2}\left[
(\dot{x}^{\underline{m}}+\partial^{\underline{m}}Y
+s_{\underline{l}}\partial^{[\underline{m}}x^{\underline{l}]})
g_{\underline{m}\underline{n}}(\dot{x}^{\underline{n}}+\partial^{\underline{n}}Y
+s_{\underline{k}}\partial^{[\underline{n}}x^{\underline{k}]})
\right.\nn\\&&\left.~
-\displaystyle\frac{1}{4}
\partial^{[\underline{m}_1}x^{\underline{m}_2]}
g_{\underline{m}_1[\underline{n}_1}g_{\underline{n}_2]\underline{m}_2}	
\partial^{[\underline{n}_1}x^{\underline{n}_2]}\right]\nn\\
L_{\rm WZ}&=&(\dot{x}^{\underline{m}_1}
	+\partial^{\underline{m}_1}Y)
C_{\underline{m}_1\underline{m}_2\underline{m}_3}\partial^{\underline{m}_2}x^{\underline{m}_3}
	+\frac{1}{6}(\partial^{\underline{m}_1}x^{\underline{m}_2})
	(\partial^{\underline{m}_3}x^{\underline{m}_4})
s_{[\underline{m}_1}C_{\underline{m}_2\underline{m}_3\underline{m}_4]}
~~~.~\label{M5Lag}
\eea

\par
\vskip 6mm
\subsection{Non-perturbative M2-brane from ${\cal M}$5-brane}\label{section:6-2}

A non-perturbative membrane action in the 11-dimensional supergravity theory is given by \cite{Bergshoeff:1987cm}
\bea
&&I~=~\displaystyle\int d^3\sigma~ L~~,~~L=L_0+L_{WZ}~,~\nn\\
&&
{\renewcommand{\arraystretch}{1.4}
	\left\{\begin{array}{ccl}
		L_0        & = & -T\sqrt{-\det \partial_\mu x^m \partial_\nu x^n g_{mn}}                                                        \\
		L_{\rm WZ} & = & \frac{T}{3!}\epsilon^{\mu\nu\rho} \partial_\mu x^{m_1}\partial_\nu x^{m_2} \partial_\rho x^{m_3} C_{m_1m_2m_3}
	\end{array}\right.}\label{M2Lagrangian}
\eea
with the spacetime index $m=0,1,\cdots,10$ and the world-volume index $\mu=0,1,2$.
The canonical coordinates are $x^m$ and
$p_m$, and the spacial world-volume coordinate derivative is
$\partial_i$ with $i=1,2$. The Hamiltonian is given by \cite{Hatsuda:2012vm} where $p_m=\partial L/\partial \dot{x}^m$
\bea
H&=& p_m\dot{x}^m-L\nn\\
&=&\lambda_0{\cal H}_\tau+\lambda^i {\cal H}_i~\nn\\
&&{\renewcommand{\arraystretch}{1.4}
	\left\{\begin{array}{ccl}
		{\cal H}_\tau   & = & \frac{1}{2}\dd_{a}{\eta}^{ab}\dd_b+
		\frac{1}{8}\dd^{a_1a_2}\eta_{a_1[b_1}{\eta}_{b_2]a_2}\dd^{b_1b_2} \\
		{\cal H}_i & = & \partial_i x^mp_m
	\end{array}\right.}\label{M2Ham}~~~.
\eea
Here $\dd_A=(\dd_a,~\dd^{ab})$ is related to
$\dd_M=(\dd_m=p_m,~\dd^{mn}=\epsilon^{ij}\partial_i x^{m_1}\partial_j x^{m_2})$
as
$\dd_A=E_A{}^M\dd_M$  for the background gauge field $E_A{}^M$.
$E_A{}^M$ includes  $g_{mn}$ and $C_{mnl}$.
The Virasoro constraint ${\cal S}^{m}=0$ in \bref{VirasoroSL4} is related to the constraint ${\cal H}_i=0$ in \bref{M2Ham} which generates  $\sigma$-diffeomorphism by multiplying the world-volume embedding operator in \bref{stwvMix} as ${\cal S}^{m}={\cal H}_i\epsilon^{ij}\partial_j x^{m}$.

We focus on the 4-dimensional subspace where the supergravity background is a representation of the SL(5) U-duality symmetry,
$E_A{}^M$ $\in$ SL(5)/SO(5).
The currents $\dd_{\underline{m}}$ and $\dd_{\underline{m}\underline{n}}$ are 4 and 6 components of 
SL(4) with  $\underline{m}=1,\cdots,4$,
which are unified into a SL(5) tensor $\dd_{mn}=(\dd_{\underline{m}},\dd_{\underline{m}\underline{n}})$ with $m=1,\cdots,5$.
The currents for a M2-brane in 4-dimensional space  \bref{M2Ham} obtained from the membrane Lagrangian \bref{M2Lagrangian} are written as
\bea
{\renewcommand{\arraystretch}{1.4}
	\left\{\begin{array}{ccl}
		{\dd}_{\underline{m}}                  & = & p_{\underline{m}}     \\
		{\dd}_{\underline{m}_1\underline{m}_2} & = &
		{\frac{1}{2}}\epsilon_{\underline{m}_1\cdots \underline{m}_4}\epsilon^{ij} \partial_i x^{\underline{m}_3}\partial_j x^{\underline{m}_4}
	\end{array}\right.}
\label{M2currents}~~~.
\eea
Commutators of \bref{M2currents} are given as
\bea
{\renewcommand{\arraystretch}{1.4}
	\left\{\begin{array}{ccl}
\left[\dd_{\underline{m}}(\sigma),\dd_{\underline{n}}(\sigma')\right]
&=&0\\
\left[\dd_{\underline{m}_1}(\sigma),\dd_{\underline{m}_2 \underline{m}_3}(\sigma')\right]
&=&2i\epsilon_{\underline{m}_1\cdots  \underline{m}_4}\epsilon^{ij}\partial_jx^{\underline{m}_4}
\partial_i\delta^{(2)}(\sigma-\sigma')\label{CA5check}\\
\left[\dd_{\underline{m}_1\underline{m}_2}(\sigma),\dd_{\underline{m}_3 \underline{m}_4}(\sigma')\right]
&=&0	\end{array}\right.}
\label{M2SL5CA}~~~.
\eea
The $p$-brane current algebras with the non-perturbative winding modes $dx^{m_1}\wedge \cdots \wedge dx^{m_p}$ are obtained similarly in
\cite{Hatsuda:2012uk}.

Now let us compare the SL(5) current algebra of the non-perturbative M2-brane \bref{CA5check}
with the one of the ${\cal M}$5-brane \bref{SL5SL4} .
The perturbative ${\cal M}$5-brane current algebra in	\bref{SL5SL4} reduces into
the non-perturbative M2-brane algebra in  \bref{CA5check}
by reducing the 5-dimensional world-volume of the
${\cal A}$5-brane into the 2-dimensional world-volume of the 
non-perturbative M2-brane as
\bea
&\partial^{\underline{m}} =\epsilon^{ij}\partial_j x^{\underline{m}}\partial_i
~~~.
&\label{stwvMix}
\eea
The operator $\partial_j x^{\underline{m}}$ is an embedding of the  membrane world-volume  to the 5-brane world-volume (where the 5-th brane coordinate is in the internal space). It has the constant form 
$\partial_j x^{\underline{m}}=\delta_j ^{\underline{m}}$ 
in the  static gauge for the ground state \cite{Hatsuda:2023dwx}.

Now we plug the world-volume projection \bref{stwvMix} into the ${\cal M}$5-brane Lagrangian \bref{M5Lag}.
The first term in the $Y=0$ gauge is given by
\bea
&&
(\dot{x}^{\underline{m}}
+s_{\underline{l}}\epsilon^{ij}\partial_j x^{\underline{m}} \partial_i x^{\underline{l}} )
g_{\underline{m}\underline{n}}
(\dot{x}^{\underline{n}}
+s_{\underline{k}}\epsilon^{i'j'}\partial_{j'} x^{\underline{n}} \partial_{i'} x^{\underline{k}} )
\nn\\
&&~~~~~=~
(\dot{x}^{\underline{a}})^2
+2s_{\underline{b}} \dot{x}_{\underline{a}}\epsilon^{ij}
\partial_j x^{\underline{a}} \partial_i x^{\underline{b}}
+(s_{\underline{b}} \epsilon^{ij}\partial_j x^{\underline{a}} 
\partial_i x^{\underline{b}} )^2  \nn\\
&&~~~~~=~
h_{00}
-2\lambda^i h_{0i}+\lambda^i\lambda^j h_{ij}
\eea
with
\bea
h_{ij}=\partial_i x^{\underline{a}}\partial_j x_{\underline{a}}=
\partial_i x^{\underline{m}}g_{\underline{m}\underline{n}}\partial_j x^{\underline{n}}
~~,~~\partial_i x^{\underline{a}}\equiv e_{\underline{m}}{}^{\underline{a}}\partial_i x^{\underline{m}}
\eea
and the membrane vielbein $\lambda^i$ and the 5-brane vielbein $s_a$ 
\bea
\lambda^i=s_{\underline{a}} \epsilon^{ij}\partial_j x^{\underline{a}}~~,~~s_{\underline{a}}=e_{\underline{a}}{}^{\underline{m}}s_{\underline{m}}~~~.
\eea
The second term is given by
\bea
&&\frac{1}{2}
(\frac{1}{2}\epsilon^{\underline{mn}}{}_{\underline{l}_1\underline{l}_2}
\epsilon^{ij}\partial_jx^{\underline{l}_1}\partial_ix^{\underline{l}_2})^2
~=~-\frac{1}{2}
(\epsilon^{ij}\partial_j x^{\underline{m}}
\partial_i x^{\underline{n}})^2
~~=~-
~{\rm det}~h_{ij}~~~,
\label{2ndterm}
\eea		
where the following relation is used in the last equality of \bref{2ndterm}
\bea
{\rm det}~h_{ij}&=&\frac{1}{2}\epsilon^{ii'}h_{ij}h_{i'j'}\epsilon^{jj'}
=\frac{1}{2}\epsilon^{ii'}
\partial_i x^{\underline{a}}\partial_j x_{\underline{a}}
\partial_{i'} x^{\underline{b}}\partial_{j'} x_{\underline{b}}
\epsilon^{jj'}=\frac{1}{2}(\epsilon^{ij}\partial_j x^{\underline{a}}
\partial_i x^{\underline{b}})^2 ~~.
\eea
We choose the following gauge of the membrane world-volume metric
\bea
\phi=-\displaystyle\frac{\sqrt{-h}h^{00}}{2}~~,~~h={\rm det}~h_{\mu\nu}~~,~~
\lambda^i=-\frac{h^{0i}}{h^{00}}~~,~~
g^2=\frac{-1}{h(h^{00})^2}~~~.\label{gaugechoicewv}
\eea
Using with the relation
\bea
{\rm det}~h_{ij}=h~h^{00}
\eea
the kinetic term $L_0$ in \bref{M5Lag} becomes
\bea
L_0
&=&\frac{1}{2}\Bigl\{
-\sqrt{-h}-\sqrt{-h}\left(
h^{00}h_{00}+2h^{0i}h_{0i}+\frac{h^{0i}h^{0j}}{(h^{00})^2}h_{ij}
\right)\Bigr\}\nn\\
&=&-\sqrt{-h}~~~.\label{NG}
\eea
This is nothing but the Nambu-Goto Lagrangian for a membrane.
The Wess-Zumino term $L_{\rm WZ}$ is obtained by using the world-volume projection \bref{stwvMix} into \bref{M5Lag} as
\bea
L_{\rm WZ}&=&\dot{x}^{\underline{m}_1}\epsilon^{ij}\partial_j
{x}^{\underline{m}_2}\partial_i x^{\underline{m}_3}C_{\underline{m}_1
\underline{m}_2\underline{m}_3}~=~\frac{1}{3!}
\epsilon^{\mu\nu\rho}
\partial_\mu{x}^{\underline{m}_1}\partial_\nu{x}^{\underline{m}_2}\partial_\rho x^{\underline{m}_3}C_{\underline{m}_1
\underline{m}_2\underline{m}_3}~~~.
\eea
Together with the Nambu-Goto term \bref{NG} 
the non-perturbative M2-brane Lagrangian is obtained from the perturbative ${\cal A}5$-brane as
\bea
I_{\rm M2}&=&\displaystyle\int d^{3}\sigma ~ L~,~L~=~L_0+L_{\rm WZ}\nn\\
L_0&=&-\sqrt{-{\rm det} ~\partial_{\mu} x^{\underline{m}}
\partial_{\nu} x^{\underline{n}}g_{\underline{m}\underline{n}}
}\nn\\
L_{\rm WZ}&=&\displaystyle\frac{1}{3!}\epsilon^{\mu\nu\rho}
\partial_{\mu}
{x}^{\underline{m}_1}\partial_{\nu}
{x}^{\underline{m}_2}
\partial_{\rho} x^{\underline{m}_3}C_{\underline{m}_1
	\underline{m}_2\underline{m}_3}~~~.
\eea
This is the expected M2-brane Lagrangian \bref{M2Lagrangian} where we set $T=1$.
\par\vskip 6mm

\section{Discussion}

In this paper we have shown how the conventional strings and membrane are obtained from ${\cal A}$-theory five-brane
with the SL(5) U-duality symmetry.

The following topics are interesting for future problems.  
\begin{enumerate}
\item{From ${\cal A}$5-brane to D-branes:
The ${\cal A}$-theory background vielbein field includes the R-R gauge fields
which couple to D-branes.
The Nambu-Goto Lagrangian will be obtained analogously to the non-perturbative  M2-brane Lagrangian as in subsection \ref{section:6-2}
with special care of the $B$-field.
The Wess-Zumino term will be obtained by adding total derivative term with the $B$-field cloud,
in such a way that the gauge transformation rule of the R-R gauge field involves the $B$-field.}
\item{From ${\cal A}$-theory branes to the non-perturbative M5- and NS5-branes:
The superstring theories admit the NS5-brane solutions which couple to the $B$-field magnetically. M-theory features the M5-brane whose U-duality symmetry is realized by the current algebra \cite{Hatsuda:2013dya}, while type IIB superstring theory contains both the NS5-brane and D5-brane related by S-duality.
These 5-brane Lagrangians are expected to be derived from ${\cal A}$5-brane and 
all such 5-branes should be connected via duality transformations.
It is interesting to clarify the structure of the 5-brane WEB including ${\cal A}$5-brane for Lagrangians analogous to the one for current algebras \cite{Hatsuda:2020buq}.}
\item{From open ${\cal A}$-theory branes to heterotic strings and type I string: 
The Lagrangians of open ${\cal A}$-theory branes \cite{Hatsuda:2021ezo}, which involve the SO(32) and E8$\times$E8 gauge groups, as well as other half-BPS branes, are of particular interest.}
\item{Quantization of ${\cal A}$- and ${\cal M}$-branes:
The main motivation for constructing the perturbative ${\cal A}$-brane Lagrangian is to facilitate a simpler quantization procedure.
Quantum effects in string theory, including winding modes of strings and branes, play a crucial role in understanding Planck-scale physics, such as the resolution of the early-universe singularity.
Quantizing ${\cal A}$-theory may provide valuable insights into a unified description of string spectra and S-matrices
 \cite{Feng:2004tg,Elvang:2009wd,Elvang:2010xn,Siegel:2020gws}.}
\item{Higher dimensional cases: 
${\cal A}$-theories in dimensions D$>$3 possess U-duality symmetry E$_{\rm D+1}$
\cite{Cremmer:1978ds,Cremmer:1979up,Bossard:2018utw,Bossard:2021jix}.
In these cases, the spacetime and world-volume dimensions become so large that they necessitate a new interpretation of the unphysical components of spacetime and world-volume.
The construction of ${\cal A}$-theory may offer a new perspective on the fundamental description of string theory.}
\end{enumerate}


\subsection*{Acknowledgments}

We are grateful to Olaf Hohm, Igor Bandos, Martin Ro\v{c}ek and Yuqi Li for the fruitful discussions. 
M.H. would like to thank Yuho Sakatani for useful discussions.
We also acknowledge the Simons Center for Geometry and Physics for its hospitality during
``The Simons Summer Workshop in Mathematics and Physics 2023 and 2024" 
where this work has been developed.
W.S. is supported by NSF award PHY-2210533.
M.H. is supported in part by 
Grant-in-Aid for Scientific Research (C), JSPS KAKENHI
Grant Numbers JP22K03603 and JP20K03604.
\printbibliography
\end{document}